\documentclass[aps,prx,showpacs,preprintnumbers,superscriptaddress,twocolumn]{revtex4-2}
\usepackage{amsmath,amssymb}
\usepackage{bm}
\usepackage{tipa}
\usepackage{upgreek}

\usepackage{tikz}

\usepackage{comment}
\usepackage{mathrsfs}
\usepackage{graphicx}
\usepackage{braket}
\usepackage{mathbbol}
\usepackage{booktabs}
\usepackage{enumitem}
\usepackage{amssymb}
\usepackage{enumitem}
\usepackage{gensymb}
\usepackage[normalem]{ulem}
\usepackage{arydshln}
\usepackage{nicematrix}
\usepackage{color}
\usepackage{graphicx}
\usepackage[colorlinks,bookmarks=true,citecolor=blue,linkcolor=red,urlcolor=blue]{hyperref}
\usepackage{hyperref}
\usepackage{cleveref}
\renewcommand{\vec}[1]{\mathbf{#1}}
\usepackage{tikz-feynman}
\makeatletter
\let\old@makecaption=\@makecaption
\usepackage{subcaption}
\let\@makecaption=\old@makecaption
\makeatother

\usepackage{relsize}

\usepackage[skins]{tcolorbox}

\newtcolorbox{myframe}[2][]{%
  enhanced,colback=white,colframe=black,coltitle=black,
  sharp corners,boxrule=0.6pt,
  attach boxed title to top left={yshift=-0.3\baselineskip-0.4pt,xshift=2mm},
  boxed title style={tile,size=minimal,left=0.5mm,right=0.5mm,
    colback=white,before upper=\strut},
  title=#2,#1
}

\renewcommand{\ket}[1]{\left. \middle| #1 \right\rangle}

\begin{document}

\title{Majorana edge reconstruction and  the $\nu=5/2$ non-Abelian  thermal Hall puzzle}
\author{Tevž Lotrič}
\affiliation{Rudolf Peierls Centre for Theoretical Physics, Parks Road, Oxford, OX1 3PU, UK}

\author{Taige Wang}
\affiliation{Department of Physics, University of California at Berkeley, Berkeley, CA 94720, USA} \affiliation{Material Science Division, Lawrence Berkeley National Laboratory, Berkeley, CA 94720,
USA}

\author{Michael P. Zaletel}
\affiliation{Department of Physics, University of California at Berkeley, Berkeley, CA 94720, USA} \affiliation{Material Science Division, Lawrence Berkeley National Laboratory, Berkeley, CA 94720, USA}

\author{Steven H. Simon}
\affiliation{Rudolf Peierls Centre for Theoretical Physics, Parks Road, Oxford, OX1 3PU, UK}

\author{S. A. Parameswaran}
\affiliation{Rudolf Peierls Centre for Theoretical Physics, Parks Road, Oxford, OX1 3PU, UK}

\begin{abstract}
Pioneering  thermal transport measurements on two-dimensional electron gases in high magnetic fields have demonstrated that the quantized Hall state  at filling factor $\nu=5/2$  has a thermal Hall conductance $\kappa$ quantized in half-integer multiples of $\kappa_0 = {\pi^2 k_B^2 T}/{3h}$. Half-integer $\kappa/\kappa_0$ is a signature of neutral Majorana edge modes, in turn linked to the presence of non-Abelian anyon excitations in the bulk.  However, the  experimentally observed value of $\kappa$ corresponds to the  `PH-Pfaffian' state, in tension with numerical studies which instead favor either the Pfaffian or the AntiPfaffian.  A variety of mechanisms have been invoked to explain this discrepancy, but have been either ruled out by further experiments or else involve fine-tuning. Building on density-matrix-renormalization group studies of physically realistic edges and analytic calculations of edge structure, we propose an alternative resolution of this puzzle involving an `edge reconstruction' solely involving the neutral Majorana sector of the theory. Such a Majorana edge reconstruction can ``screen'' a Pfaffian or AntiPfaffian bulk, so that transport signatures become indistinguishable from those of the PH-Pfaffian. We argue that this physically natural scenario is  consistent with experiment. 
\end{abstract}
\maketitle

\section{Introduction} 

The nature of the
 fractional quantized Hall (FQH) plateau at Landau level (LL) filling factor $\nu=5/2$ has been a matter of intense debate since its discovery~\cite{52_state_discovery_PhysRevLett.59.1776}. Several distinct incompressible phases of matter have been proposed to explain the experiments, with the preponderance of numerical evidence~\cite{morf_1998_PhysRevLett.80.1505,RezayiHaldane2000,Peterson2008,Wan2008,feiguin_2009_PhysRevB.79.115322,Zhao2011,simon_2011_truncated_LL_PhysRevLett.106.116801,some_more_pert_LL_mixing_PhysRevB.91.205404,29_Zaletel_2015,pakrouski_2015_PhysRevX.5.021004,rezayi_2017b_PhysRevLett.119.026801,Yutushui2020,Rezayi2021,Mila,Henderson} indicating that the most likely candidates are the Moore-Read Pfaffian (Pf)~\cite{58_MOORE1991362} or its particle-hole conjugate, the AntiPfaffian (aPf)~\cite{for_pf_aPf_disorder_PhysRevLett.99.236807,26_Levin_PhysRevLett.99.236806} phase of matter. However, a groundbreaking series of experiments~\cite{48_Banerjee_2018,with_noise_experiment,50_Dutta_2022,25_paul2023topological} have instead consistently favored an interpretation in terms of a third phase,  the PH-Pfaffian (PHPf)~\cite{DamSon_Dirac_fermions_PhysRevX.5.031027}.   All three possibilities host quasiparticle excitations with non-Abelian statistics,  meaning that they are candidates for topological information processing~\cite{NayakReview}, and hence {are} particularly exciting for fundamental and practical reasons. 
 
 In light of this, the tension between theory and experiment is unsatisfactory. Although it is possible that the theories fail to capture some essential feature favoring PHPf in experiments, the situation is in stark contrast to the remarkable agreement at several other filling factors whose quasiparticles have Abelian fractional statistics~\cite{first_heat_other_fract_Banerjee_2017}. An early attempted reconciliation, advanced by one of us, {relied} on the observation that  experiments only probe physics at the {edge}, and {suggested that incomplete equilibration of a Majorana edge mode would explain the measured thermal conductance}~\cite{simon_interpetation_of_thermal_conductance_PhysRevB.97.121406,03_PhysRevLett.124.126801,FeldmanComment,SimonReply}. Although initially promising, this explanation --- which only applies to the aPf --- has been significantly challenged, particularly given further experiments~\cite{with_noise_experiment,50_Dutta_2022,25_paul2023topological}. An alternative approach focused instead on the bulk, and explored a scenario wherein the effect of interactions on disorder-induced puddles of Pf and aPf~\cite{18_Wang_2018,puddles_2_PhysRevLett.121.026801,puddles_1_PhysRevB.97.165124}  could lead to PHPf physics at long wavelengths. However, the necessary conditions seem quite fine-tuned~\cite{18_Wang_2018}, making it less plausible as a realistic explanation of experiments.

\begin{figure}
    \centering
    \includegraphics[width=\columnwidth]{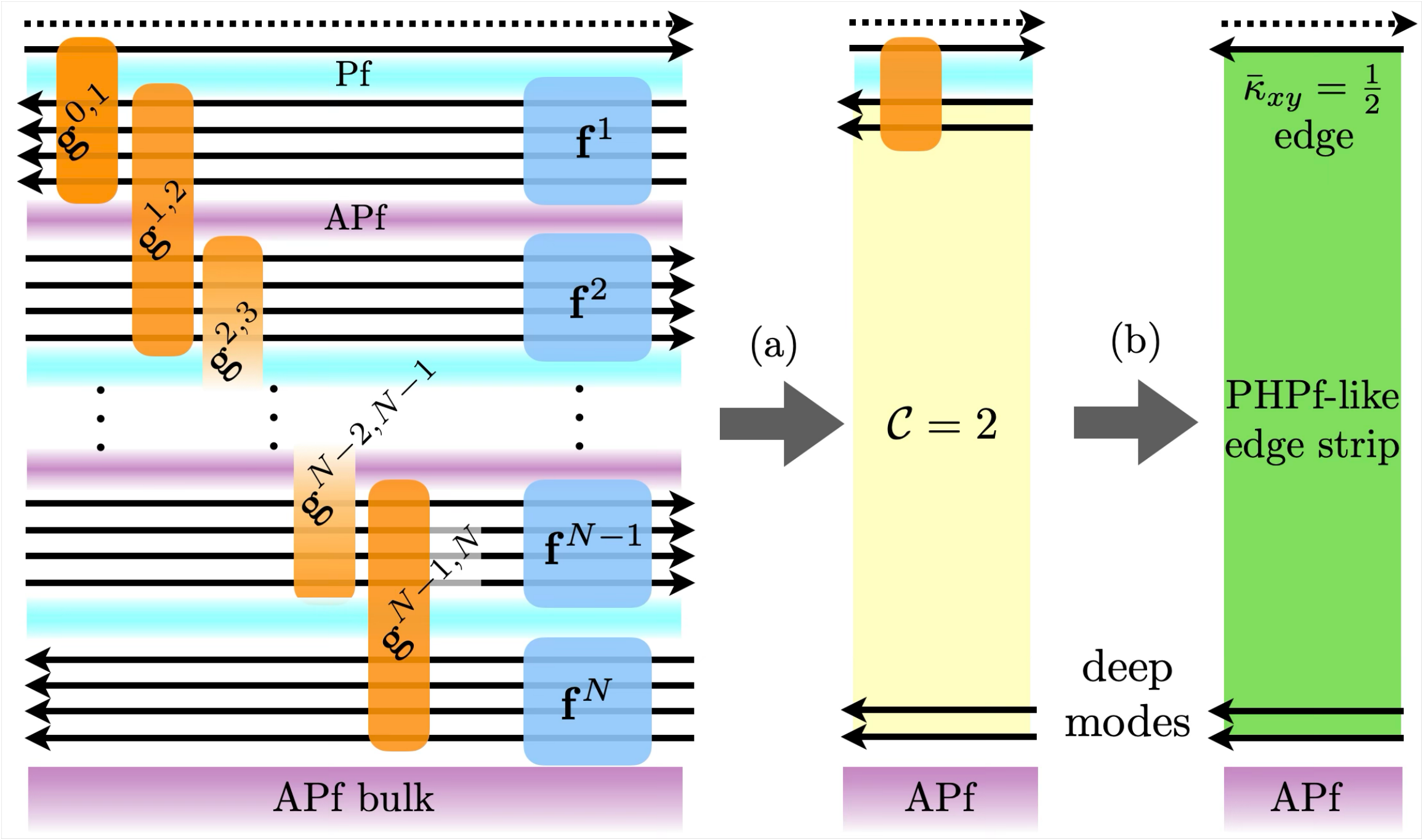}
    \caption{\textbf{Majorana Edge Reconstruction.}  Majorana mode quadruplets (solid arrows) bound to Pf/aPf domain walls near the edge (a) first  gap out `pairwise' to produce a $\mathcal{C}=2$ strip, whose edge modes in turn (b) couple with the charge (dashed) and Majorana (solid) mode  at the physical boundary to produce an effective $\bar{\kappa}_{xy}=1/2$ edge and a pair of `deep modes' separated by a mesoscopic strip of PHPf. Edge transport measurements are insensitive to the deep modes that do not equilibrate with physical edge modes due to their large separation from them.}
    \label{fig:modes_sketch}
\end{figure}

Here we propose a new and natural resolution to the conflict between theoretical predictions and experimental observations at $\nu=5/2$.  Our approach builds on the numerical observation, via density-matrix renormalization group (DMRG) studies, of a novel form of reconstruction at the edge of a half-filled LL. For experimentally realistic confining potentials, the incompressible QH liquid {exhibits very strong density oscillations extending over  $L_e\sim 40\ell_B$ into the bulk (recall that the  magnetic length $\ell_B = (\hbar/eB)^{1/2}$ is the natural scale for a LL in magnetic field $B$). Surprisingly, the chirality of the entanglement spectrum oscillates in tandem with the electron density at the location of the entanglement cut. This finding suggests the formation of  alternating stripes of Pf and aPf near the edge.} {A similar  structure for the entire {bulk} sample was proposed in Ref.~\cite{WanYangStripes}, but in the present scenario the stripe pattern is not a property of a stable bulk phase but only exists in a finite region near the edge.}  Each of the series of Pf/aPf interfaces that results from this reconstruction hosts four Majorana modes, with successive quadruplets propagating in alternating directions.  Over a wide range of physically reasonable parameters, we show analytically that backscattering between adjacent quadruplets and between the outermost quadruplet and the original edge modes results in an effective PH-Pfaffian edge. We find that this PHPf-like edge {is} nevertheless  consistent with {\it either}  a bulk Pf or bulk aPf phase, due to the emergence of a pair of chiral modes deep in the bulk that combine with it to satisfy the mandates of the bulk-boundary correspondence, {as illustrated for the edge structure of an aPf bulk in Fig~\ref{fig:modes_sketch}}. We show that edge thermal  measurements are insensitive to the deep modes:  the emergent mesoscopic  length scale $L_e$ ``screens'' the {bulk} Pf {or} aPf, so that edge experiments  diagnose PHPf behaviour. Remarkably, this picture, although blending aspects of both previous proposed resolutions, in contrast to them does not require fine-tuning and --- via the emergent mesoscopic scale $L_e$ ---  gives a concrete physical mechanism for the lack of equilibration. We also demonstrate that reasonable assumptions on bulk-edge coupling mechanisms and interaction scales suffice to explain the observed temperature scaling of the measurements,  further supporting the Majorana edge reconstruction scenario.

In the remainder, we review the physics of the $\nu=5/2$ edge in Sec.~\ref{sec:reviewedge}, and then present our  numerical results that motivate the edge reconstruction scenario in Sec.~\ref{sec:DMRG}. We then provide an analytically tractable theory for the reconstruction in Sec.~\ref{sec:hybridization}. We then use this to rationalize various features of the numerical data in Sec.~\ref{sec:revisitDMRG}, before providing an analysis of the thermal conductance measurements in light of the reconstruction in Sec.~\ref{sec:edgeeq}. We close in Sec.~\ref{sec:conclusions} by discussing the implications of our results for future interferometric probes of non-Abelian statistics.

\section{The $\nu=5/2$ Edge \label{sec:reviewedge}} 
We begin by summarizing the different possible  $\nu=5/2$ edge structures. Throughout, we will ignore the two filled LLs, and focus on the half-filled level, which contributes $\sigma_{xy} = \bar{\sigma}_{xy} \sigma_0$ and $\kappa_{xy} = \bar{\kappa}_{xy} \kappa_0$ to the Hall and thermal Hall conductance.  (This is equivalent to the physical edge between the half-filled level and the $\nu=2$ integer QH state.) Here we have introduced the  natural units $\sigma_0 = e^2/h$ and $\kappa_0 = \pi^2 k_B T/3h$ for the Hall and thermal Hall conductances, so that we can work with the dimensionless quantities $\bar{\sigma}_{xy}$ and $\bar{\kappa}_{xy}$ henceforth.  The Hall conductance is fixed to $\bar{\sigma}_{xy} =1/2$; in the edge-state picture, this corresponds to a single chiral  charge-$\frac{1}{2}$ boson propagating ``downstream'', whose Lagrangian
\begin{equation}
    \mathcal{L}_c[\phi] = -\frac{2}{4\pi}(\partial_x\phi)(\partial_t\phi +v_\phi \partial_x \phi) +\ldots \label{eq:chargebase}
\end{equation}
is of the Luttinger liquid form, where `$\ldots$' denotes higher-order terms that are irrelevant in the renormalization group (RG)  sense. 
In contrast, there are different possibilities for $\bar{\kappa}_{xy}$; since the charge mode gives a fixed downstream contribution $\bar{\kappa}^c_{xy} =1$, edges with different $\bar{\kappa}_{xy}$ are distinguished by the presence of $|n|$  chiral neutral Majorana modes that may propagate either {{all}} downstream (parallel to the charge mode, $n>0$) or {{all}} upstream (anti-parallel to it, $n<0$). These may be  described by 
\begin{equation}
  \mathcal{L}_{\text{M}}^{n}[\{\psi_l\}] = \sum_{l,l'=1}^{|n|} \psi_{l}(i\partial_t\delta_{ll'}  + h_{ll'}^{\text{sign}(n)}(\vec{v}, \vec{f}))\psi_{l'} +\ldots 
\end{equation}
where $h^{\pm1}_{ll'}(\vec{v}, \vec{f}) = i [ \pm v_{ll'}\partial_x  +  f_{ll'}]$, $v_{ll'} = v_l\delta_{ll'}\geq0$ is a diagonal matrix of mode speeds, $f_{ll'}=-f_{l'l}$ is an antisymmetric matrix of inter-mode couplings,
and as before `$\ldots$' denotes less RG-relevant terms (at least quartic in Majoranas or quadratic in derivatives). Since all $|n|$ modes are chiral, $\mathbf{f}$ cannot gap them and therefore $\mathcal{L}_{\text{M}}^{n}$ contributes  $\bar{\sigma}_{xy} =0$ and (assuming full equilibration) $\bar{\kappa}_{xy} = {n}/{2}$.

The simplest case is an edge with a single ``downstream'' Majorana mode, with Lagrangian   $\mathcal{L}_{\text{Pf}} = \mathcal{L}_c + \mathcal{L}_{\text{M}}^{1}$ and  hence  $\bar{\kappa}_{xy} = \bar\kappa_{xy}^c + \frac{1}{2} = \frac{3}{2}$.
This edge theory corresponds to a bulk in the non-Abelian QH phase described by the Moore-Read ``Pfaffian'' wavefunction~\cite{58_MOORE1991362}. 

To motivate the other possible edge structures, we observe that under a particle-hole transformation  $\mathcal{P}$ within a LL, we have $(\bar{\sigma}_{xy}, \bar{\kappa}_{xy}) \overset{\mathcal{P}}{\rightarrow} (1-\bar{\sigma}_{xy}, 1-\bar{\kappa}_{xy})$. For the Pfaffian, 
$(\bar{\sigma}_{xy}, \bar{\kappa}_{xy}) = (1/2, 3/2) \overset{\mathcal{P}}{\rightarrow} (1/2, -1/2)$. Since the transformed value of $\bar{\sigma}_{xy}$ remains consistent with a downstream charge mode with $\bar\kappa_{xy}^c = 1$, we conclude that the particle-hole transformed edge consists of a single charge mode and three \textit{upstream} neutral Majorana modes. Thus, after a particle-hole transformation, the edge is described by the Lagrangian $\mathcal{L}_{\text{aPf}} = \mathcal{L}_c + \mathcal{L}_{\text{M}}^{-3}$; evidently, this is distinct from $\mathcal{L}_{\text{Pf}}$ and corresponds to a distinct bulk phase of matter, captured by the so-called ``AntiPfaffian'' state~\cite{for_pf_aPf_disorder_PhysRevLett.99.236807,26_Levin_PhysRevLett.99.236806} (which may be obtained from the Pfaffian by applying $\mathcal{P}$ in the bulk).

\begin{figure*}[tbp]
\includegraphics[width=0.95\textwidth]{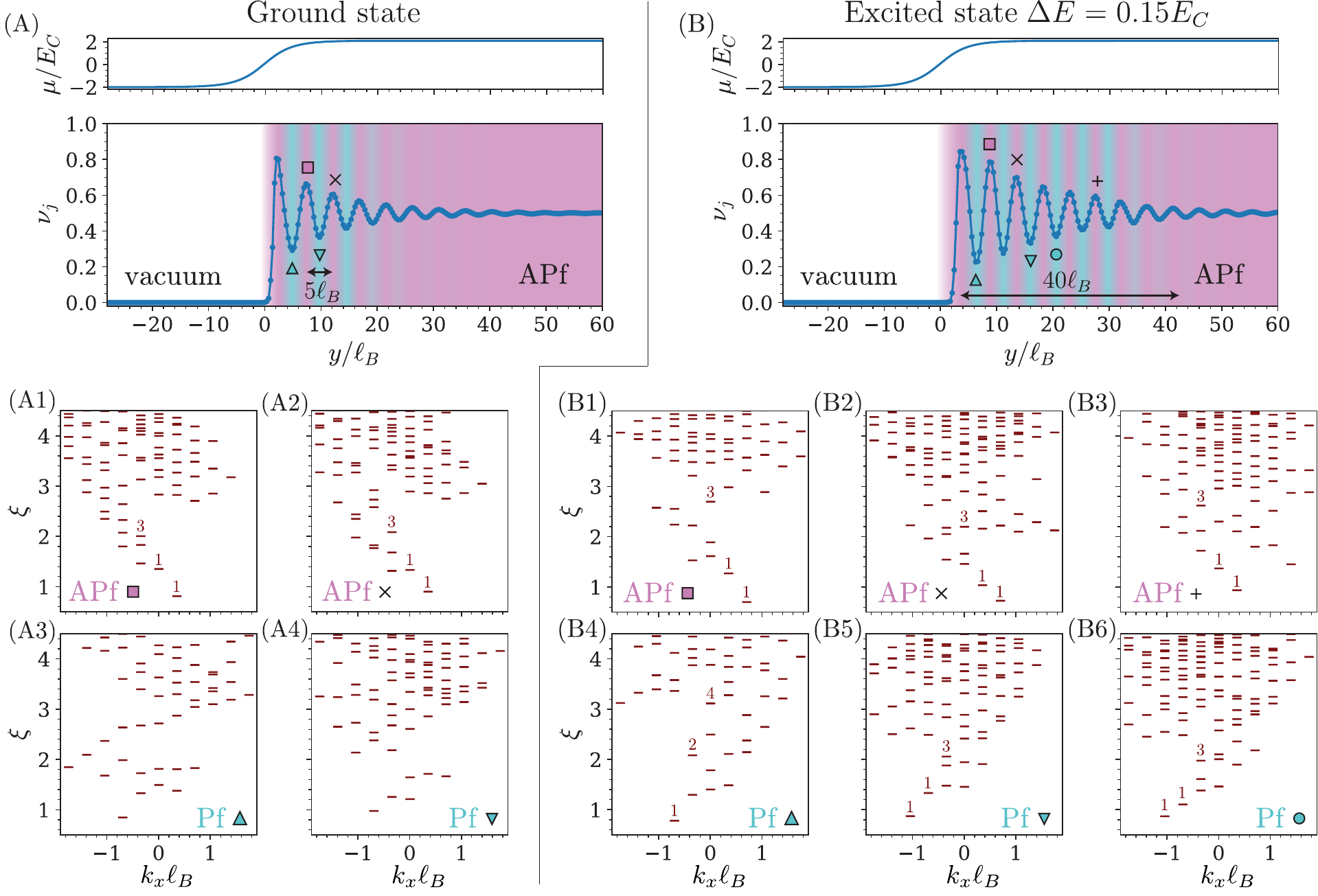}
\caption{\textbf{Edge Reconstruction in DMRG.}
(A)~Ground state   
with background chemical potential as in top panel.
(B)~A low-lying excited state in the momentum sector $\delta k_x = -15.5 \times 2\pi/L_x$ relative to the ground state. 
\textit{Top panels:} orbital occupation $\nu_j$ and local chemical potential $\mu(y) = - U(y)$. The confining potential induces charge‑density oscillations that penetrate roughly $L_{e}\approx40\ell_{B}$ into the bulk with spatial period $5\ell_{B}$. Regions are shaded blue (Pf) or red (APf) according to the chirality inferred from the orbital entanglement spectrum (OES).  \textit{Bottom panels:} OES at representative cuts. Anti‑chiral level counting $(1,1,3,\dots)$ or $(1,2,4,\dots)$ appears in (A1–A2) and (B1–B3), matching the APf edge theory.  Chiral counting in (B4–B6) follows the same counting, consistent with Pf; the ground‑state segments (A3–A4) are likewise chiral, though their level counting remains inconclusive. Results were obtained on a cylinder of circumference $L_{x}=18\ell_{B}$ with gate distance $d=5\ell_{B}$, chemical potentials $\mu_{L}=-2E_C$ and $\mu_{R}$ at the middle of the bulk gap, and bond dimension $\chi=2048$.
}
\label{fig:SharpDMRG}
\end{figure*}

The final possibility is an edge {\it invariant} under $\mathcal{P}$, which requires that $(\bar{\sigma}_{xy}, \bar{\kappa}_{xy}) = (1/2, 1/2)$. It is straightforward to see that such an edge must consist of a downstream charge mode and a {\it single} upstream Majorana mode, $\mathcal{L}_{\text{PH}} = \mathcal{L}_c + \mathcal{L}_{\text{M}}^{-1 }$, which corresponds to the ``PH-symmetric Pfaffian'' phase introduced in Ref.~\onlinecite{DamSon_Dirac_fermions_PhysRevX.5.031027}.

Armed with this background, we now describe the central issue to be addressed.  The thermal transport experiments~\cite{48_Banerjee_2018,25_paul2023topological} {as well as 
edge noise measurements~\cite{50_Dutta_2022} }are consistent with  PHPf.   In contrast, numerical studies of the bulk phase --- which reliably match experiments at other filling factors --- consistently support Pf or aPf, but never  PHPf~\cite{morf_1998_PhysRevLett.80.1505,RezayiHaldane2000,Peterson2008,Wan2008,feiguin_2009_PhysRevB.79.115322,Zhao2011,simon_2011_truncated_LL_PhysRevLett.106.116801,some_more_pert_LL_mixing_PhysRevB.91.205404,29_Zaletel_2015,pakrouski_2015_PhysRevX.5.021004,rezayi_2017b_PhysRevLett.119.026801,Yutushui2020,Rezayi2021,Mila,Henderson}. As noted above, proposed resolutions involving poorly equilibrated edges or bulk disorder are unsatisfactory, so for guidance we turn to a numerical study of the structure of a physically realistic edge.

\section{Edge reconstruction via DMRG \label{sec:DMRG}}
Edge reconstruction is a ubiquitous response of quantum Hall edges to soft confinement.  
Whenever the external potential varies smoothly on the scale of the magnetic length $\ell_B$, the system can lower its Coulomb energy by breaking the edge into several narrow incompressible strips~\cite{ChamonEdge,ReimannEdge99,WanEdge2003,Karzig2012Edge,WangEdge2013,KhannaEdge21}.  
This ``composite edge’’ supports a larger set of gapless modes than would be inferred from the bulk topological order alone; however, the total electrical Hall conductance $\bar\sigma_{xy}$ and the thermal Hall conductance $\bar\kappa_{xy}$ remain pinned as long as {all the modes from all} sub‑edges have fully equilibrated~\cite{KaneFisherThermalEdge}. The equilibration scale is set by the coupling between sub-edges, in turn controlled by microscopic details such as the strength of interactions and disorder and the spatial separation of the charged and neutral modes, which is typically a few magnetic lengths at most.

At filling factor $\nu = 5/2$ the reconstruction becomes more intricate. Because the half-filled second Landau level supports several nearly degenerate incompressible phases, one can imagine a scenario in which different phases are locally realized as one moves inward  from the sample edge.
Boundaries between these phases generate subtle electrostatic effects. For instance, a Pfaffian–anti-Pfaffian domain wall can carry a finite dipole moment, which Coulomb interactions and the external confining potential can selectively stabilize, thereby reshaping the edge profile \cite{38_Park_2014,12_PhysRevLett.125.146802}.

To evaluate the possibility of such reconstructions, we study an edge between the $\nu = 5/2$ and $\nu = 2$ states under realistic electrostatic conditions using  the ``defect‑DMRG’’ method of Ref.~\onlinecite{ZaletelMongPollman}. An infinitely long cylinder with circumference $L_x$ is partitioned such that far to the left of a wide central region the state relaxes to the $\nu = 2$ integer quantum Hall vacuum, whereas far to the right it relaxes to the $\nu = 5/2$ aPf phase, realizing the edge of interest in the large central region.

The precise model is as follows. To facilitate convergence, long‑range Coulomb interactions are screened by metallic gates at $z = \pm d$, giving $V_d(q) = \tanh(qd)\,2\pi/q$, in units of $E_C = e^2/(4\pi\varepsilon\ell_B)$ units.  
The bottom gate is kept grounded, while the top gate is split into a left and right at voltages $V_L$ and $V_R$. Solving the Laplace equation for this split-gate geometry produces a smoothed  step‑like potential $\mu(y) = e \phi(y)$ at the 2DEG (see Fig.~\ref{fig:SharpDMRG} and~\cite{Supplement}),  resulting in Hamiltonian 
\begin{equation}
H = \frac12 \int d\mathbf r\,d\mathbf r'\; :\rho(\mathbf r)V_d(\mathbf r - \mathbf r')\rho(\mathbf r') :
       - \int d\mathbf r\;\mu(\mathbf r)\rho(\mathbf r),
\end{equation}
where $\rho$ is the electron density and $: \, \, :$ denotes normal ordering. 
The interactions and potential are projected into the $N = 1$ LL, so henceforth the filling $\nu$ is measured relative to the inert $\nu = 2$ background.

On the aPf side, $V_R$ is adjusted to pin the chemical potential to the middle of aPf gap,
$\mu_R = \tfrac42\bigl(\Delta_{+e/4} - \Delta_{-e/4}\bigr)$, where $\Delta_{\pm e/4}$ are the energies to add or remove an $e/4$ quasiparticle, which we independently measured using excited-state DMRG.  
On the vacuum side we merely require the chemical potential to lie within the (large) $\nu = 2$ cyclotron gap $-\hbar\omega_c < \mu_L < 0$ (here $\mu$ is measured relative to $\nu = 2 + \epsilon$). This leaves a range of admissible $V_L$, which can be used to tune the sharpness of the edge. Detailed scans confirm that the qualitative reconstruction is insensitive to the precise choice of $V_L$~\cite{Supplement}.

To obtain the ground state in the presence of the edge using DMRG, we first obtain the bulk ground states at $\nu = 0$ and $\nu = 1/2$ filling of an $N = 1$ LL; the latter case produces the aPf.  
We then consider a matrix product state (MPS) ansatz in which the MPS tensors far to the left are taken from the $\nu=0$ ground state,  and far to the right are taken from the $\nu = 1/2$ aPf ground state, while $N_s$ tensors in an intermediate window between them are variationally optimized to approximate the ground state. ~\cite{ZaletelMongPollman}. To speed convergence, we initialize the intermediate window with a block of vacuum tensors followed by aPf tensors.  Changing the location at which the initial ansatz switches from aPf to vacuum changes the total charge; subsequent DMRG sweeps preserve this charge, allowing us to sweep through the $E(Q)$ landscape.

Fig~\ref{fig:SharpDMRG}~(A) shows the ground‑state charge density profile with $L_x = 18\ell_B$, gate spacing $d = 5\ell_B$, chemical potential $\mu_L = -2E_C$, and bond dimension $\chi = 2048$. Pronounced oscillations of period $\lambda\sim 5\ell_B$ persist over a remarkably long decay length $L_e \sim 40\ell_B$. 
To reveal the local topological structure underlying the density oscillation, we examine the orbital entanglement spectrum (OES)~\cite{li2008entanglement} for cuts at different locations.

Deep in the aPf bulk the orbital entanglement spectrum (OES) displays the expected anti-chiral counting of the aPf state. Moving toward the edge, panels (A1–A4) of Fig. \ref{fig:SharpDMRG} track the OES at successive density extrema. At the maxima (A1–A2) the low-lying levels follow the sequences $(1,1,3,\dots)$ and $(1,2,4,\dots)$ that signal the $\mathbb 1$ and $\chi$ anyon sectors of the aPf phase.~\cite{li2008entanglement} Strikingly, at the minima (A3–A4) the chirality appears reversed,  though the counting is ambiguous. In the remainder of the paper we will argue these oscillations may be interpreted as alternating Pf and aPf stripes, whose edges hybridize at low energy.  Because the Pf and aPf have different Hall viscocities, an interface between them  is expected to carry a  quantized dipole moment  in proportion to the  mismatch.~\cite{HaldaneViscocity} Thus alternating Pf and aPf stripes are expected to produce charge-density oscillations at the same  $\lambda \sim 5\ell_B$ period.~\cite{38_Park_2014,12_PhysRevLett.125.146802}

In the ground state data just discussed, however, there is no obvious identification of the reversed OES as the Pf. Further evidence for such an interpretation emerges from low-lying excited states with a momentum boost relative to the ground state. In Fig.~\ref{fig:SharpDMRG}(B), we show one  such state in the $\delta k_x \approx -15.5 \times 2\pi/L_x$ sector and energy $\delta E \approx 0.15 E_C$ above the ground state. For technical reasons related to our DMRG implementation, the total charge also differs by $\delta Q = 1.5 e$. 
The density oscillations look very similar to those of the ground state, but with somewhat increased amplitude. 
Most remarkably, the low-lying OES now shows not just reversed chirality, but striking resemblance to either the Pf or aPf OES in perfect correspondence with the density oscillations. We refer to the Supplement~\cite{Supplement} for the OES of various other states in the low-energy manifold.

To summarize: in both the ground state and momentum-boosted state the OES oscillates in tandem with the charge density, but only in the large momentum state does the OES clearly correspond to alternating stripes of aPf / Pf.
To explain why, we propose a two-step mechanism. At a higher energy scale, the edge breaks into narrow stripes that alternate between Pf and aPf phases, whose interface carries gapless Majorana modes. At a lower energy scale, these neighboring Majorana modes overlap and hybridize, which adds counter-propagating branches to the OES and hides the simple Pf counting. When a finite momentum boost is applied, we will show that Pauli exclusion turns off the low-$k$ part of the hybridization. As a result, the Pf counting reappears in the low-$k$ part of the boosted excited-state OES.

Although the behavior described above is robust across a broad range of $V_L$, $V_R$, and $d$, it is not universal. As shown in the Supplement~\cite{Supplement}, much softer edges obtained by setting $\mu_L = -0.1 E_C$ also display  density oscillations (albeit of reduced magnitude), yet the OES follows Pf counting throughout the $L_e \approx 40 \ell_B$ edge region before crossing over to aPf in the bulk. This behavior fits within the same two step framework: Pf and aPf stripes first nucleate at the edge, and complete hybridization of their interface modes can yield an extended edge region with either Pf, aPf, or PHPf order, depending on microscopic details. Because the hybridization depends sensitively on edge sharpness, interaction range, and excitation content, we observe several distinct patterns under different confinement profiles~\cite{Supplement}. The next sections develop an analytic framework for this hybridization, clarifying the range of possible edge structures, which we can then use to interpret the DMRG data.

\section{Majorana Edge Hybridization\label{sec:hybridization}}
By comparing the  Lagrangians $\mathcal{L}_{\text{Pf}}$, $\mathcal{L}_{\text{aPf}}$, it is clear that an isolated interface between Pf and aPf  is generically described by four co-propagating Majorana modes, $\mathcal{L}^{\pm 4}_{\text{M}}$, with a negative or positive sign for Pf/aPf or aPf/Pf interface, respectively. The width of an isolated Pf/aPf interface $w$ is of the same order as the size $\lambda/2$ of the strips in our system \cite{12_PhysRevLett.125.146802}, so we generically expect the Majorana modes from adjacent interfaces (which {counter-}propagate) to interact and gap each other out --- an effect that is challenging to fully capture via DMRG due to the small energy scales involved. To address the reconstruction, we therefore take the stripe pattern from DMRG as an input, and study the hybridization of the resulting low-energy modes analytically. 

We model a striped edge parallel to the $x$-axis by a series of $N$ strips alternating between Pf  and aPf extending from the vacuum at $y=0$ to the bulk uniform QH liquid at $y = L_e = (N/2)\lambda$. (Two strips comprise a full oscillation, and for Fig.~\ref{fig:SharpDMRG}(B), $N \sim 16, \lambda \sim 5\ell_B$.) We  take the strip adjacent to vacuum to be Pf, so that the bulk is aPf for $N$ odd, and Pf for $N$ even (our results can readily be adapted to when the first strip is aPf, but we make this choice for simplity of exposition). Moving to a Hamiltonian description and ignoring disorder, the Majoranas are described by $H=H_0 + H_{4N} + H_{c} + H_{c0}$, where
\begin{align}
H_0 &= \sum_{k>0} \psi_0(-k)(v_0 k) \psi_0(k), \,\,\, \text{and} 
\\
H_{4N} &= \sum_{k>0} 
\sum_{\alpha=1}^N\sum_{l,l'=1}^4 \psi_{\alpha,l}(-k)[  (-1)^{\alpha}v_{ll'} k + i f^\alpha_{ll'}] \psi_{\alpha, l}(k)\nonumber
\end{align}
 respectively represent the single chiral mode at the $y=0$ vacuum-Pf interface and the $N$ quadruplets of modes at the Pf/aPf interfaces, and 
\begin{align} \label{eq:gs_def}
H_{{c}} &= i\sum_{k>0}\sum_{\alpha, \alpha'=1}^{N} \sum_{l,l'=1}^4 \psi_{\alpha,l}(-k) g^{\alpha,\alpha'}_{l,l'}\psi_{\alpha',l'}(k) \,\,\, \text{and}  \\
H_{c0} &= i\sum_{k>0}\sum_{\alpha=1}^N\sum_{l=1}^4 g^{0,\alpha}_{l}[\psi_0(-k) \psi_{\alpha,l}(k) - \psi_{\alpha,l}(-k) \psi_0(k)]\nonumber
\end{align} 
describe the couplings between the different Pf/aPf interfaces and between the interfaces and the vaccum edge, with $g^{\alpha,\alpha'}_{l,l'} = -g^{\alpha',\alpha}_{l',l}$. 
Next, we assume that all couplings except those between adjacent interfaces are negligible, so that  only $g^{\alpha, \alpha\pm 1}_{ll'}$ and $g^{0,1}_{l}$ are non-zero (this is physically reasonable for local interactions; see Fig~\ref{fig:modes_sketch}). We then have $H = \sum_{k>0}\boldsymbol{{\psi}}_{-k}\mathbf{M}_k\boldsymbol{{\psi}}_{k}$, where $\boldsymbol{{\psi}}_{k} =(\psi_0(k), \psi_{1,1}(k), \psi_{1,2}(k), \ldots, \psi_{N,4}(k))$ and
\begin{widetext}
\renewcommand\arraystretch{1.1}
\begin{equation}
   \mathbf{M}_k = \left[\begin{array}{c|ccccc} v_0 k & i\mathbf{g}^{0,1} & \mathbf{0} & \ldots &\ldots &\ldots\\
   \hline
    -i(\mathbf{g}^{0,1})^T & -\mathbf{v}^1 k + i \mathbf{f}^1 & i \mathbf{g}^{1,2} & \mathbf{0} &\ldots&\ldots\\
    \mathbf{0} & -i (\mathbf{g}^{1,2})^T  & \mathbf{v}^2 k + i \mathbf{f}^2 & i \mathbf{g}^{2,3} &\mathbf{0} &\ldots\\
     & \mathbf{0}  & -i (\mathbf{g}^{2,3})^T & -\mathbf{v}^3k + i \mathbf{f}^3 & \ldots &\ldots \\ 
    \vdots &\vdots & \vdots &\vdots &\ddots & i \mathbf{g}^{N-1,N}
    \\\vdots &\vdots & \vdots &\vdots & -i (\mathbf{g}^{N-1,N})^T &(-1)^N \mathbf{v}^N k +i \mathbf{f}^N
    \end{array}  \right].\label{eq:bigmatrixH}
\end{equation}
\end{widetext}
Note that $H$ is purely quadratic in Majoranas and we ignore higher order terms as these are RG subleading relative to the terms we keep.

At this stage, we could  in principle simply diagonalize $H$ (numerically, if necessary) to determine  the fate of the edge; however, since there are no clear symmetries and $H$ has a very large number ($\sim 16N$) of independent parameters, this offers limited insight into the physics. One approach, that we turn to momentarily, is to simply choose parameters randomly, possibly with some additional structure e.g. reflecting the spatial falloff of couplings. To motivate our approach and to build  intuition, we consider an initial toy model with a hierarchy of scales of the form 
$|g^{\alpha,\alpha+1}_{(1\text{ or } 2),(3 \text{ or }4)}| \sim | f^{\alpha}_{(1\text{ or } 2),(1 \text{ or }2)}|\sim | f^{\alpha}_{(3\text{ or } 4),(3 \text{ or }4)}|,$
with all other couplings in $\mathbf{f}$ and $\mathbf{g}$ negligible relative to these.  In this simplified limit, is easiest to first ignore $H_{c0}$ and work out what happens to $H_{4N} +H_c$ (the large lower-right block in \eqref{eq:bigmatrixH}): the modes $3,4$ for each interface $\alpha\in\{1,2,\ldots N-1\}$ gap out against the counterpropagating modes $1,2$ from interface $\alpha+1$,  leaving gapless the pair of upstream modes $1,2$  from $\alpha=1$ and the pair of modes $3,4$  from $\alpha=N$ as shown in Fig.~\ref{fig:modes_sketch} (note that these are downstream/upstream for $N$ even/odd). Crucially, these two pairs of modes are separated by the mesoscopic scale $L_e$; thus, when we now add back in  $H_{c0}$, one of the modes in the first pair gaps out the downstream mode from $H_0$, leaving a single upstream mode. In contrast, the ``deep'' modes remain gapless and decoupled from the edge. Evidently, {\it independent of the bulk phase} the physics near the edge is that of a charge mode and one upstream Majorana, so that it is described by $\mathcal{L}_c + \mathcal{L}_{\text{M}}^{-1} = \mathcal{L}_{\text{PH}}$, the PH-Pfaffian edge theory. The bulk-boundary correspondence required by the Pf/aPf phase in the far interior is restored by the two ``deep'' modes, with $\mathcal{L}_{\text{M}}^{\pm2}$, which propagate in precisely the direction such that they combine with the  PH-like {\it physical} edge to yield $\mathcal{\mathcal{L}_{\text{Pf/aPf}}}$ for the composite edge.

To see why this result is robust, observe that $H_{4N} + H_c$ can be viewed as a Bogoliubov-de Gennes Hamiltonian written in its Majorana representation; since there is no global symmetry, this lies in Class D of the  ``tenfold way'' classification. The relevant integer topological invariant $\mathcal{C}$, counts the net number of chiral Majorana modes remaining near the $\alpha=0$ interface beyond those required by the Pf/aPf itself and can be related to a Chern number in a periodic system (which requires $N$ even). The first step of our calculation above establishes that  our toy model has $\mathcal{C}=2$ (in a convention where the Majorana sectors of the Pf/aPf have $\mathcal{C}_0=-1,3$ and $|\mathcal{C}_0+\mathcal{C}|$ Majorana modes are localized close to the edge); the second {step} then follows upon coupling $H_{4N} + H_c$ to $H_0$. Since $\mathcal{C}$ is an integer, it cannot change continuously, and hence is stable to small changes in parameters. Thus, the reconstruction represents a {\it phase} of the composite edge and hence is not fine-tuned.
 
 We can generalize the above construction by considering {\it random}  couplings between the modes (but still translationally-invariant parallel to the edge), chosen via
 \begin{equation}
f^{\alpha}_{ll'} = \mathcal{N}[0, R^{|l-l'|-1}],\,\,\,\, g^{\alpha,\alpha+1}_{ll'} = \mathcal{N}[0, R^{|l-l'+s-3|}],\label{eq:randomparametrization}
 \end{equation}
 where $\mathcal{N}[\mu,\sigma^2]$ denotes a Gaussian random variable with mean $\mu$ and variance $\sigma^2$ (sampled independently for each matrix element), and the exponent of $R\in [0,1]$ captures the effect of the spatial positions of modes on their couplings.  The `offset' $s$ controls the spatial overlap of the modes, so that mode $1$ from interface $\alpha+1$ is closest to mode $(4-s)$ of interface $\alpha$, see Table~\ref{fig:chern_probs}. 
 As flagged above, the width of an isolated interface is similar to the size of the strips in our system, making it unclear what values of $s$ and $R$ apply. This leads us to consider all possibilities.
 The scale hierarchy chosen in the simplified discussion above corresponds to taking $R\ll 1$, and $s=1$, but we can consider the problem more generally by drawing random instances of $H_{4N}+H_c$ with matrix elements distributed according to Eq.~\eqref{eq:randomparametrization}, and computing $\mathcal{C}$ for each. Determining  the relative probability of phases with different $\mathcal{C}$ under such sampling gives a rough sense of their extent in parameter space. (Note that the assumption of independently distributed random couplings may not be strictly correct for {\it physical} edge Hamiltonians, and so these probabilities should not be interpreted literally but more in the spirit of random matrix theory, as lending qualitative insight.) The results (Table~\ref{fig:chern_probs}) reveal that there is always a finite window of $\mathcal{C}=2$,  which is required for the physical edge to emulate the PH-Pfaffian without fine-tuning. The phase structure is relatively insensitive to $R$; indeed, the finite  $\mathcal{C}=2$ window persists even for $R=1$, when $s$ plays no role and there is no structure beyond the nearest-neighbour form in \eqref{eq:bigmatrixH}.  

Above, we introduced randomness to sample the parameter space of a disorder-free system. Separately, we have also studied a problem with quenched disorder in the  inter-mode couplings (i.e, we write \eqref{eq:bigmatrixH} in position ($x$) space with spatially varying random  $\mathbf{f}^\alpha(x), \mathbf{g}^{\alpha,\alpha'}(x)$), and verified that here too there is a finite window where $H_{4N}+H_{c}$ hosts a pair of chiral modes at the physical edge and the deep interior and hence effectively has $\mathcal{C}=2$~\cite{Supplement}.

\begin{table}[]
\resizebox{0.9\columnwidth}{!}
{
\begin{tabular}{cc|ccc|}
\cline{3-5}
\multicolumn{1}{l}{}       & \multicolumn{1}{l|}{} & \multicolumn{3}{c|}{Probability of $\mathcal{C}=2$ Phase}           \\ \hline
\multicolumn{2}{|c|}{Mode Offset}                  & \multicolumn{1}{c|}{$R=0$}  & \multicolumn{1}{c|}{$R=0.5$} & $R=1$  \\ \hline
\multicolumn{1}{|c}{$s=0$} & \raisebox{-6pt}{\includegraphics[height=16pt]{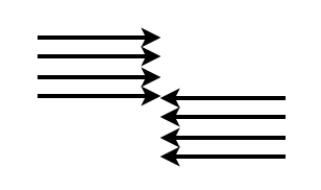}}                    & \multicolumn{1}{c|}{$0.30$} & \multicolumn{1}{c|}{$0.38$}  &        \\ \cline{1-4}
\multicolumn{1}{|c}{$s=1$} & \raisebox{-6pt}{\includegraphics[height=16pt]{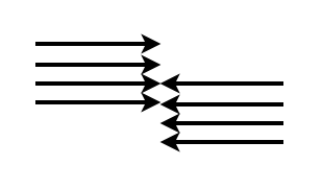}}                    & \multicolumn{1}{c|}{$0.81$} & \multicolumn{1}{c|}{$0.45$}  & $0.37$ \\ \cline{1-4}
\multicolumn{1}{|c}{$s=2$} & \raisebox{-6pt}{\includegraphics[height=16pt]{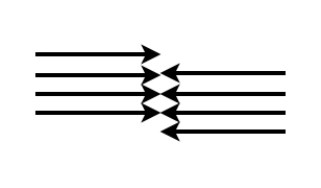}}                    & \multicolumn{1}{c|}{$0.35$} & \multicolumn{1}{c|}{$0.36$}  &        \\ \hline
\end{tabular}
}
\caption{\textbf{Stability of Reconstruction.}  Probabilities correspond to the frequency of $\mathcal{C}=2$ for different $R$ and $s$ in Eq.~\ref{eq:randomparametrization}. The  $\mathcal{C}=2$ phase appears with a finite probability for all parametrizations. Statistical uncertainties are $\sim 0.01$.  The simplified case is one of the instances with $R=0, s=1$. For $s$ negative, the distribution is similar to $s=0$.  \label{fig:chern_probs}}
\end{table}

\section{Interpretation of DMRG results\label{sec:revisitDMRG}}
With this theoretical picture in mind, we now return to the interpretation of the DMRG results. First recall that for smoother edges ($\mu_L = -0.1 E_C$), DMRG still finds density  oscillations, but the OES is consistent with the formation of a large $L_e \sim 40 \ell_B$ stripe of Pfaffian at the edge. 
Such behavior implies that the four Majorana modes bound to successive Pf--aPf interfaces have strongly hybridized into a pair of counter-propagating channels, thereby realizing the $\mathcal{C}=4$ edge configuration introduced above.
Indeed, we must emphasize  that our proposal does not \emph{require} the formation of a $\mathcal{C}=2$ PHPf boundary, it simply allows for it to be generic; consistent with this, the random matrix results show a probability for $\mathcal{C}=2$ which is large but less than 1. 
We do not currently have a microscopic understanding of how the edge sharpness relates to the  $\mathbf{f}, \mathbf{g}$ parameters of the effective  model, though this would be useful to explore in future work in order understand how experiments might tune between different regimes.  

Second, we return to our observation that when DMRG is used to target excited states in a boosted $k_x$ sector, the OES shows  clear oscillations between Pf and aPf, while the ground state does not.
To understand this, consider a much simpler, but analogous, scenario in which a strip of $p + ip$ superconductor forms in a trivial $\nu = 0$ background, resulting in left / right moving chiral Majorana modes $\psi_{L/R}$ at the boundaries of the strip. 
In the absence of hybridization between them, an entanglement cut in the middle of the $p + ip$ strip would look like that of $p + i p$ SC, e.g. right-handed chiral. 
Hybridization between the edges modifies the entanglement, resulting in an entanglement spectrum with  left and right moving modes (of different velocity) with zero net chirality. However, we will show that in states with boosted momentum, this hybridization entanglement is removed from the low-lying entanglement spectrum: the low-lying spectrum is thus expected to look like the $p + i p$ SC in the boosted-momentum state. 
An analogous calculation applies to the  multi-channel case.

To make this calculation concrete, consider left and right moving Majorana modes, $\psi_L(k), \psi_R(k)$, satisfying $\{ \psi_{\mu}(k), \psi_{\nu}(k') \} = \delta_{\mu \nu} \delta_{k+k'}$. A model for their hybridization is
\begin{align}
H = \frac{1}{2} \sum_k &\left[\psi_R(-k)(vk)\psi_R(k) + \psi_L(-k)(-vk) \psi_L({k})\right. \nonumber \\
&\left.+2 i g \psi_R({-k}) \psi_L(k) + \cdots \right]    
\end{align}
When $g=0$, the vacuum $\ket{0}_L \ket{0}_R$ is defined by the property $\psi_R(k) \ket{0}_R = \psi_L({-k}) \ket{0}_L = 0$ for $k > 0$. Thus $\psi_R({-k}), \psi_L(k)$ for $k > 0$  act like creation operators on top of this vacuum.

Following an analogous calculation in Ref.~\cite{QiKatsuraLudwig}, the hybridized ground state takes the form
\begin{equation}
\ket{g}= \prod_{k>0} \left[\sqrt{1 - p_k} + i \sqrt{p_k} \psi_R(-k)\psi_L(k)\right]
\ket{0}_L \ket{0}_R ,
\label{eq:majorana_ES}
\end{equation}
where $\ket{0}_{L/R}$ is the vacuum in the absence of hybridization and  
$p_k = \tfrac12 - \tfrac{v k}{2|g|} + \mathcal{O}(k^{3})$.  
Eq.~\eqref{eq:majorana_ES} is already in Schmidt form, with single-particle
entanglement energies determined by  
$e^{-\epsilon_k}= p_k/(1-p_k)$, giving  
$\epsilon_k = 2 v k/|g| + \cdots$.

This is not the entirety of the entanglement, however, as we must remember that the ``vacuum''
$\ket{0}_L\ket{0}_R$ is a $p + i p$ SC, which itself has a chiral entanglement spectrum $\epsilon_k^{(0)} = - v_B k$. Keeping track of the signs, its entanglement chirality is opposite to the hybridization entanglement just discussed. 
The full entanglement spectrum therefore contains two branches:  
the hybridized branch $\epsilon_k$ arising from the coupling between the
left and right modes, and the intrinsic branch $\epsilon_k^{(0)}$ inherited
from the $p + i p$ bulk. 
The hybridized branch $\epsilon_k$ can obscure the counting of the intrinsic branch $\epsilon_k^{(0)}$ when the two sets of levels overlap.

Now instead consider the state with momentum $\Delta K$ larger than the ground state. 
For a monotonic dispersion relation, this is obtained by filling the lowest $k$-state excitations up to some $k_\ast$, $\Delta K = \sum_{k < k_\ast} k$.
The resulting lowest-energy state is verified to be
\begin{equation}
\begin{split}
\ket{g, \Delta K} = \prod_{k>k_\ast}  (\sqrt{1 - p_k} + i \sqrt{p_k} \psi_R({-k}) \psi_L({k}) )\\ \times \prod_{k<k_\ast} \psi_R({-k})\ket{0}_L \ket{0}_R.
\end{split}
\end{equation}
The key observation is that the modes $k < k_\ast$ no longer contribute  entanglement, since there is no superposition in these sectors. This can alternatively be understood by noting that in the boosted state, \emph{both} the L and R $k$-modes are occupied for $k < k_\ast$, so their hybridization is Pauli blocked. 
The lowest entanglement levels $\epsilon_k$ for $k < k_\ast$ are thus removed from the entanglement spectrum (equivalently,  $\epsilon_k \to \infty$ for $k < k_\ast$). 

Boosting to large momentum thus removes the low-lying entanglement that would otherwise come from hybridization, effectively recovering the un-hybridized result. This  mechanism generalizes to the multi-channel case relevant to the Pf-aPf domains. In the boosted $\Delta K > 0$ states, the low-$k$ $\psi_{R,\alpha}(k)$ modes (where $\alpha$ now runs over wires and flavors) become filled, so that their hybridization with the L modes is Pauli blocked. We conjecture that this mechanism results in the difference in OES shown in Fig.\ref{fig:SharpDMRG}A, B, where the boosted state shows an OES that oscillates between Pf and aPf, as if the edges are unhybridized.

\section{Equilibration and Edge Transport\label{sec:edgeeq}} We have established that for a range of parameters, there is a mesoscopic separation of the ``topologically mandated'' Pf/aPf edge into a PH-Pfaffian like physical edge described by $\mathcal{L}_c[\phi] + \mathcal{L}_M^{-1}[\chi]$ and a pair of deep chiral neutral modes $\mathcal{L}_M^{\pm2}\left[\psi_1,\psi_2\right]$. In order for this structure to explain experiment, it is crucial that these two spatially separated modes remain decoupled even in the presence of generic interactions $\mathcal{L}_{\text{int}}$ and/or disorder $\mathcal{L}_{\text{dis}}$.  The most relevant couplings between $\mathcal{L}_c[\phi] + \mathcal{L}_M^{-1}[\chi]$ and $\mathcal{L}_M^{\pm2} [\psi_1,\psi_2]$ in the RG sense are terms $\sim \chi\psi_{1,2}$. But first, as $\psi_{1,2}$ are close together, it is natural to expect a term $im\psi_1\psi_2 \subset \mathcal{L}_\text{int}$. Due to the separation of interfaces, we expect this term to have a larger prefactor than any $\chi\psi_{1,2}$ term in $\mathcal{L}_\text{int}+\mathcal{L}_\text{dis}$; therefore,  we first eliminate $\psi_1 \psi_2$ by rotating to a new set of decoupled Majoranas $\tilde{\psi}_{1,2}$, but at momenta $\pm k_0$, with $k_0 \sim\mathcal{O}(\ell^{-1}_B)$ on dimensional grounds. The residual couplings of the form $\chi \tilde{\psi}_{1,2}$ can then be treated perturbatively. Already at the bare level, these are suppressed by two effects: (1)  the small wavefunction overlap between the different edge modes and (2) the momentum mismatch between $\tilde{\psi}_{1,2}$ and the single Majorana mode $\chi$ which must remain at $k=0$. The latter suppresses both interactions and disorder: in experiments, we expect the disorder potential is correlated over $\sim 100~\text{nm}\sim 10 \ell_B\gg k_0^{-1}$, making it too smooth to effectively supply the needed momentum boost. Thus, although   $\chi\tilde\psi_{1,2}$ has scaling dimension $\Delta=1$, and is hence RG relevant in both the clean and disordered cases~\cite{07_PhysRevLett.125.016801}, the RG flow for experimentally reasonable length scales and temperatures cannot overcome the two bare suppression effects, and so such terms cannot effectively equilibrate the physical edge with the deep modes \cite{Supplement}. All other allowed terms coupling $\phi,\chi$ to $\psi_{1,2}$ necessarily have  higher scaling dimensions while still suffering from the same suppression effects so they can be ruled out as sources of equilibration as well. We argue in the Supplement that these effects also prevent coupling of $\psi_{1,2}$ to the floating-point contacts used in experiments.{This complete decoupling of the deep modes is crucial; note that, while lack of equilibration between modes is generally argued to {\it increase} $\bar\kappa_{xy}$, this argument no longer applies when some subset of modes, beyond just being out of equilibrium with other modes, also further decouple from the contacts~\cite{Supplement}. Taken together, this means that the deep modes will not contribute to thermal Hall measurements.}

We are thus left with a PH-Pfaffian edge theory $\mathcal{L}_\text{PH}\left[\phi,\chi\right]$, whose modes must be brought into equilibrium to match experiments. Previous work~\cite{01_PhysRevB.107.245301,07_PhysRevLett.125.016801} has argued that the most relevant coupling between the modes is via the highly RG-irrelevant ($\Delta=3$) operator  $\left(\partial_x\phi\right) \chi i\partial_x\chi$. We estimate that the bare coupling necessary for this to produce equilibration consistent  with the experiments is $\sim 10^3$ times larger than reasonable dimensional estimates, so this is unlikely to be the dominant mechanism~\cite{Supplement}. Furthermore, the equilibration length resulting from such an operator is $\sim T^{2-2\Delta}\sim T^{-4}$ which diverges rapidly at low temperatures and would likely drive an increase in $\bar{\kappa}_{xy}$ at low $T$. While this was seen in early experiments~\cite{48_Banerjee_2018}, it has not been reproduced in subseqeuent ones~\cite{with_noise_experiment,50_Dutta_2022,25_paul2023topological}.
These factors lead us to consider an alternative mechanism. In the presence of a Majorana zero mode $\gamma$ localized at an impurity close to the edge, the modes may be coupled via the $\Delta=\frac32$ operator $\left(\partial_x\phi\right) \chi\gamma$. This is RG marginal in the disordered problem and is likely strong enough to equilibrate the modes in experiment. It also suggests a milder scaling $\sim T^{-1}$ of the equilibration length, consistent with the lack of low-$T$ growth in $\bar{\kappa}_{xy}$. {Note that \cite{25_paul2023topological} measures the upstream and downstream contributions to $\bar\kappa_{xy}$ separately and thus does  not rely on PHPf equilibration. But equilibration is necessary to match direct measurements of $\bar\kappa_{xy}$ \cite{48_Banerjee_2018,50_Dutta_2022,with_noise_experiment}.} We defer further discussion of the equilibration length calculations to the Supplement~\cite{Supplement}.

\section{Concluding Remarks\label{sec:conclusions}}  {While our calculation was performed considering an interface between $\nu=5/2$ and $\nu=2$ with either a Pf or aPf bulk, the physics is essentially identical if we consider the interface between $\nu=5/2$ and $\nu=3$ --- a PHPf edge can form next to the integer state, which is measured by any transport experiment, and two additional states are left deep in the bulk, decoupled from edge measurements.   Thus our theory can be brought into agreement with all  prior experiments~\cite{48_Banerjee_2018,with_noise_experiment,50_Dutta_2022,25_paul2023topological}, as long as we assume a sufficiently steep confining potential. In summary, we have demonstrated that it is possible to separate the bulk topological physics from the observed edge, and that for a range of reasonable parameters this separation can be made consistent with experiment.

How might this new edge structure  affect Fabry-Perot interferometry measurements (e.g., Ref.~\onlinecite{WillettInterferometry2023})?  Since the charge modes nearest the edge must be partially reflected from the point contacts, it is reasonable to assume that the deep modes are fully reflected, thus forming a disconnected dot in the middle of the cavity. In this case, one should see interference of the PH-Pfaffian edge modes, irrespective of the ``true'' bulk order. Similar arguments apply to other possible edge structures for other choices of inter-Majorana coupling or edge confinement scale. The central lesson remains that the link between bulk and edge in the non-Abelian $\nu=5/2$ QH liquid is more subtle than previously imagined, and is a richer problem than dictated by topological considerations alone.}

\begin{acknowledgements} 
We are indebted to Roger Mong and Frank Pollmann for development of the DMRG code used in this work.
This work was supported by the European Research Council under the European Union
Horizon 2020 Research and Innovation Programme
via Grant Agreements No. 804213-TMCS (SAP), and by the UKRI under a  Frontier Research Guarantee (for an ERC Consolidator Grant) EP/Z002419/1 (SAP)  and  EPSRC Grants EP/S020527/1 and EP/X030881/1 (SHS). MZ and TW were supported by the U.S. Department of Energy, Office of Science, Office of Basic Energy Sciences, Materials Sciences and Engineering Division, under Contract No. DE-AC02-05CH11231, within the Theory of Materials program (KC2301). SAP acknowledges the hospitality of the Max Planck Institute of Complex Systems, Dresden, where part of this work was completed with support from a Gutzwiller Award. This research used the Lawrencium computational cluster provided by the Lawrence Berkeley National Laboratory (Supported by the U.S. Department of Energy, Office of Basic Energy Sciences under Contract No. DE-AC02-05CH11231)
\end{acknowledgements}

\bibliography{references}

\onecolumngrid

\newpage
\clearpage

\begin{center}
    {\large\bf Supplementary Material for:``{Majorana edge reconstruction and  the $\nu=5/2$ non-Abelian  thermal Hall puzzle}''
}

\vspace*{10pt}

\end{center}

\setcounter{section}{0}
\setcounter{equation}{0}
\setcounter{figure}{0}
\setcounter{table}{0}
\setcounter{page}{1}
\makeatletter
\renewcommand{\theequation}{S\arabic{equation}}
\renewcommand{\thefigure}{S\arabic{figure}}
\renewcommand{\thetable}{S\arabic{table}}

\section{DMRG calculation}

\subsection{Setup}

\textit{Step 1—Bulk ground states.}
We work on an infinitely long cylinder of circumference \(L_{x}=18\ell_{B}\).
Metallic screening planes at \(z=\pm d\) with \(d=5\ell_{B}\) render the Coulomb interaction short-ranged, 
\begin{equation}
V_{d}(q)=\frac{2\pi}{q}\tanh(qd),
\end{equation}
in units of \(E_{C}=e^{2}/4\pi\varepsilon\ell_{B}\).
Results for \(d=10\ell_{B}\) are qualitatively identical; we therefore quote data for \(d=5\ell_{B}\).

The Hamiltonian takes the form
\begin{equation}
H=\frac12\int d^{2}\mathbf r d^{2}\mathbf r'
\rho(\mathbf r) V_{d}(\mathbf r-\mathbf r') \rho(\mathbf r').
\end{equation}
where $\rho(\mathbf r)$ is the density operator projected to the \(N=1\) Landau level (LL). Neglecting LL mixing, the Moore–Read Pfaffian (Pf) and its particle–hole conjugate, the anti-Pfaffian (aPf) are exactly degenerate; we choose the aPf sector, the phase expected to prevail once LL mixing is reinstated.

\textit{Step 2—Chemical-potential calibration.}
The ``defect''-DMRG protocol~\cite{31_PhysRevLett.110.236801,taige_gap_2024} isolates individual anyons in a wide window whose left and right boundaries are pinned to topological sectors \([l,r]\).  
For the Pfaffian family
\([0110,1010]\) traps a charge \(+e/4\) quasiparticle and  
\([0110,0101]\) a charge \(-e/4\) quasihole, yielding  
\begin{equation}
E_{+e/4}=0.54E_{C},\qquad
E_{-e/4}=-0.51E_{C}
\end{equation}
for \(L_{x}=18\ell_{B}\) and \(d=5\ell_{B}\).
The shared Madelung term cancels in the difference, fixing the mid-gap chemical potential to  
\begin{equation}
\mu_{\mathrm{aPf}}=\frac{4}{2}\bigl(E_{+e/4}-E_{-e/4}\bigr)=2.10E_{C}.
\end{equation}

\textit{Step 3—Edge-reconstruction setup.}
A split top gate held at \(V_{L}\neq V_{R}\) and a grounded bottom gate (\(V_{B}=0\)) generate the step-like potential  
\begin{equation}
U(y)=\frac{V_{L}}{2}
      +\frac{V_{R}-V_{L}}{\pi}
       \arctan(e^{\pi y/2d}),
\end{equation}
where \(y=0\) sits at the interface and the smoothness is controlled by \(d\).
Target chemical potentials \(\mu_{L}\) (vacuum) and \(\mu_{R}\) (aPf) fix the gate voltages to \(V_{L/R} =  - 2 \mu_{L/R}\), with \(\mu_{R} = - 2 E_C\) chosen inside the cyclotron gap.

Beginning with the homogeneous MPS tensors obtained in Step 1, we construct an inhomogeneous ansatz composed of three regions: a semi-infinite vacuum lead on the left, a semi-infinite aPf lead on the right, and a window of \(N_{s}\) variational tensors in between.  The window is seeded with \(N_{\mathrm{vac}}\) vacuum tensors followed by \(N_{\mathrm{aPf}}=N_{s}-N_{\mathrm{vac}}\) aPf tensors.

Ideally one would minimize the energy in every sector labeled by the conserved quantum numbers—the charge \(Q\) and the total momentum \(K_x\) in the window. At finite bond dimension \(\chi\), however, the variational freedom afforded by a given seed characterized by \(N_{\mathrm{vac}}\) is limited by modest fluctuations in \((Q,K_x)\). We therefore sweep \(N_{\mathrm{vac}}\); each increment \(N_{\mathrm{vac}} \to N_{\mathrm{vac}}+1\) transfers charge \(\Delta Q=-e/2\) and produces a correlated shift in \(K_x\).  In this way every charge sector is accessible, but for a fixed \(Q\) the set of reachable momentum \(K_x\) is still restricted at finite \(\chi\). For the relatively sharp confining potential considered here, the true ground state always falls within the accessible \(K_x\) set.  By varying \(N_{\mathrm{vac}}\)—and the allowed \((Q,K_x)\) for each seed—we sweep each ansatz to convergence and retain the complete low-energy manifold.

\begin{figure}[t]
  \centering
  \includegraphics[width=0.65\linewidth]{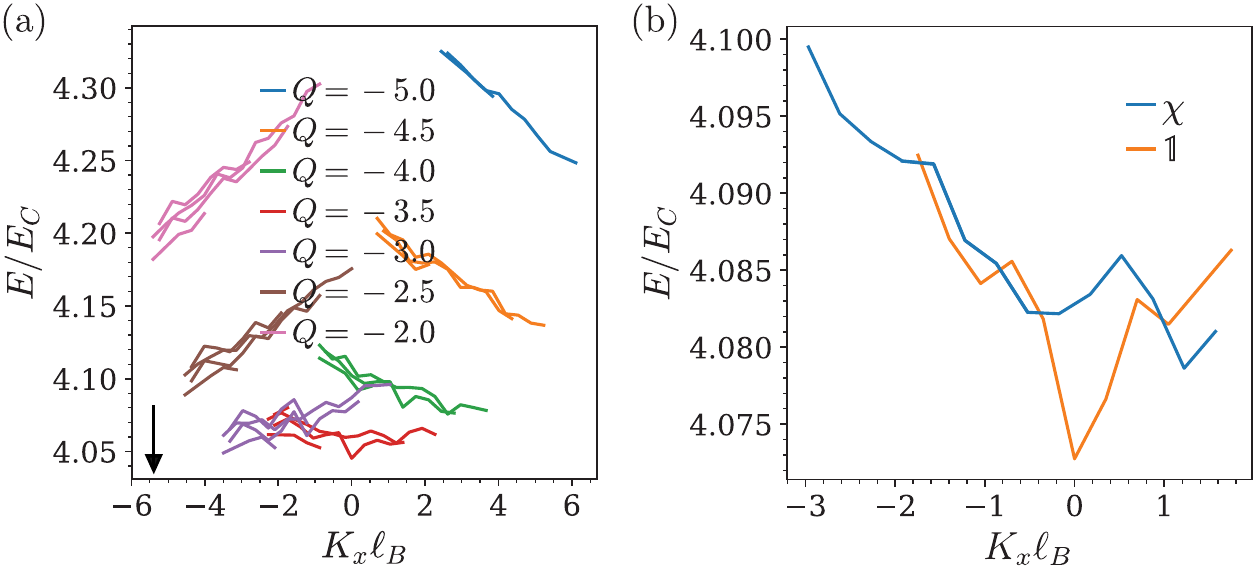}
  \caption{Edge energy \(E\) as a function of total momentum \(K_x\).
(a) Comprehensive \(Q\) survey obtained with \(N_{s}=104\) and \(\chi=512\); colors distinguish different charge sectors. {The black arrow indicates the momentum $K_x\approx-15.5\times2\pi/L_x$ of the state in Fig.~\ref{fig:SharpDMRG}(B)}. The collapse is not perfect due to limited \(N_{s}\) and \(\chi\).
(b) Enlarged view of the optimal sector \(Q=-3.5e\), calculated with \(N_{s}=360\) and \(\chi=2048\) where we see perfect collapse.
Two topological branches are resolved: the trivial anyon sector \(\mathbb 1\) (orange) and the neutral-fermion sector \(\chi\) (blue).
Energies are quoted in units of \(E_{C}\). {The slight difference in energies between (a) and (b) is a result of using different $N_s$ and $\chi$.}
}
  \label{fig:QKscan}
\end{figure}

\begin{figure*}[htbp]
  \centering
  \includegraphics[width=\textwidth]{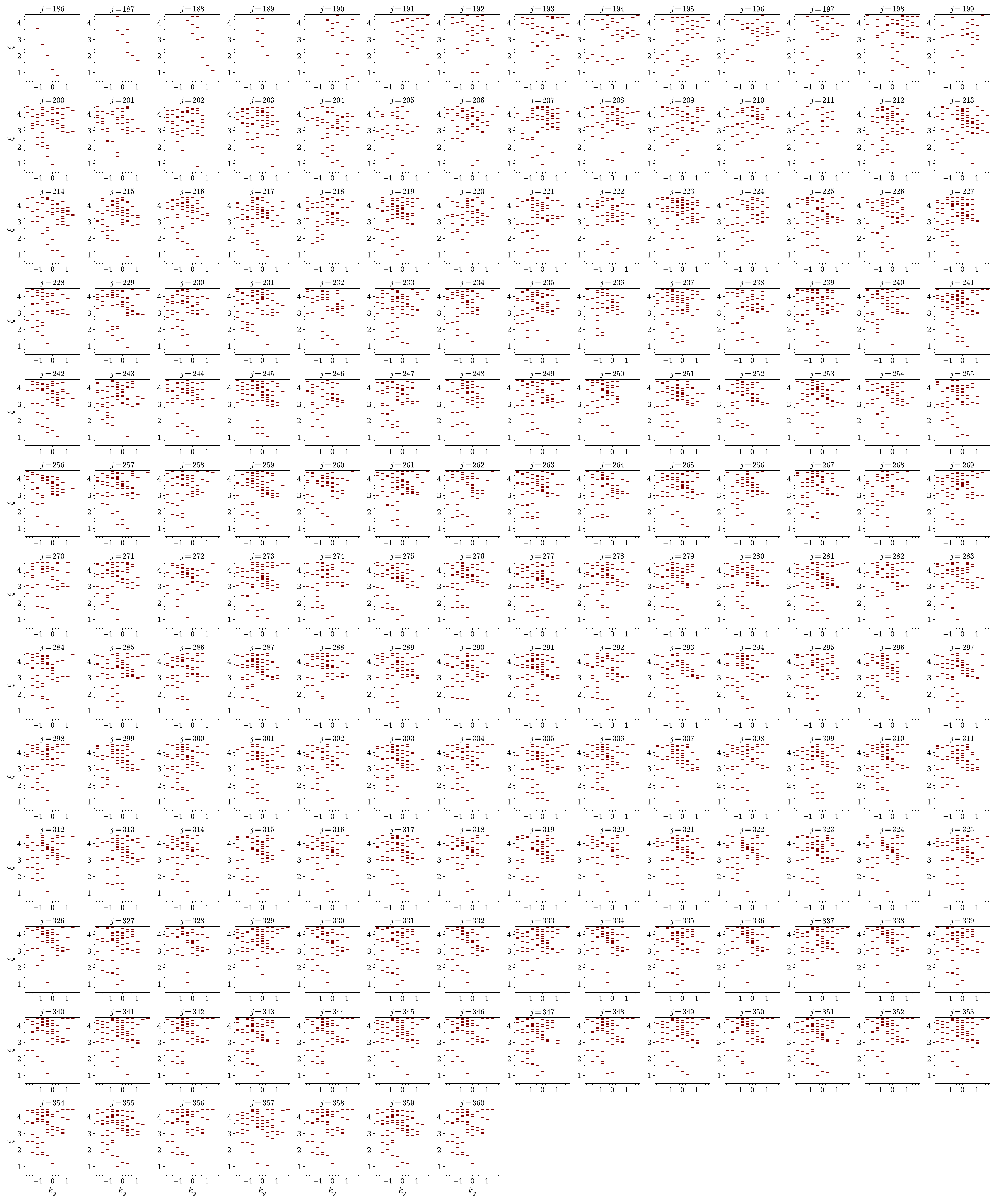}
  \caption{Entanglement spectra at every bond \(j\) of the \(\chi=2048\) ground state shown in main-text Fig~2(A). Anti-chiral aPf counting at density maxima alternates with chiral counting at density minima.}
  \label{fig:OESFullgs}
\end{figure*}

\begin{figure*}[htbp]
  \centering
  \includegraphics[width=\textwidth]{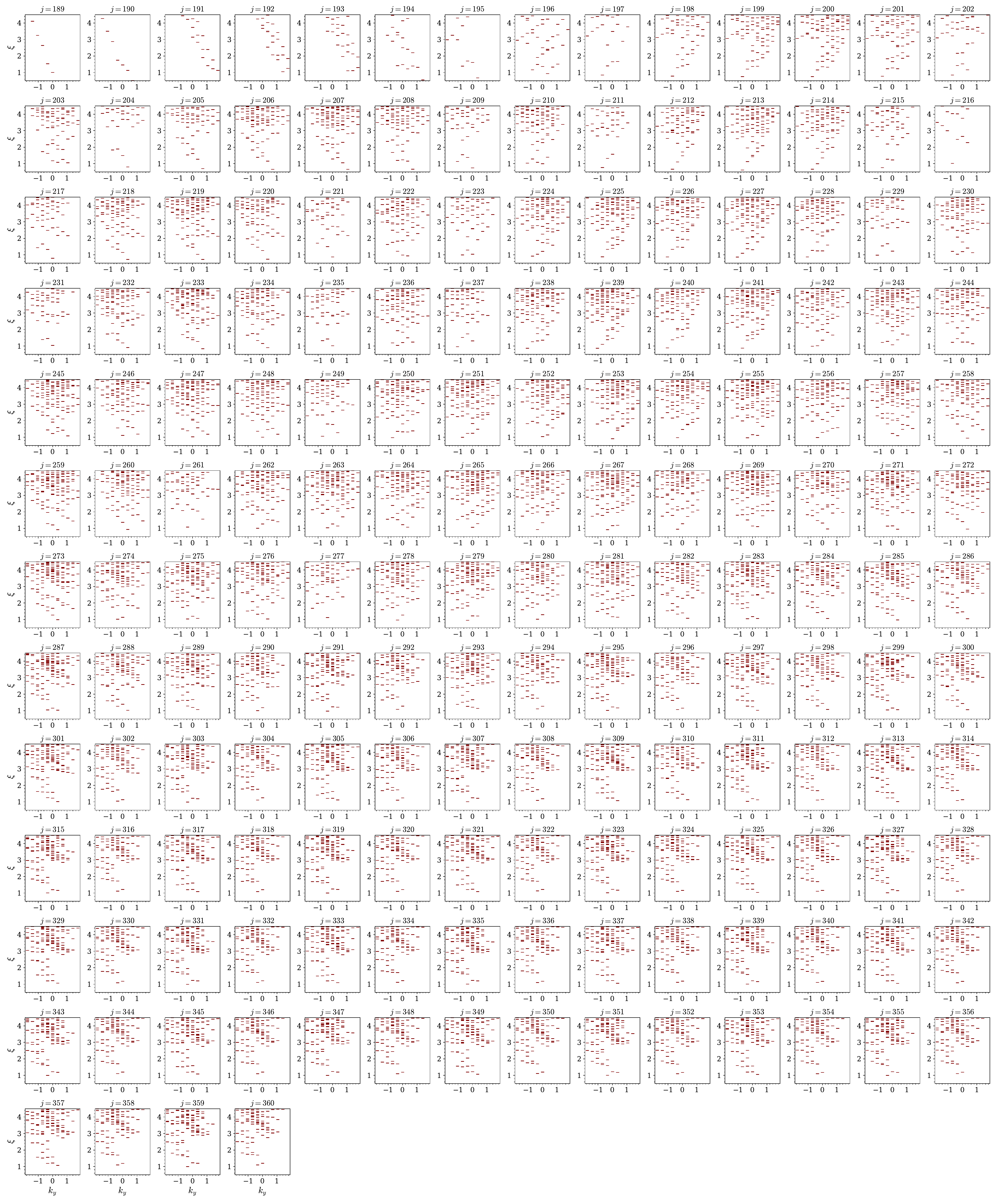}
  \caption{Entanglement spectra at every bond \(j\) of the \(\chi=2048\) excited state shown in main-text Fig~2(B). Density minima shows clear Pf counting.}
  \label{fig:OESFullext}
\end{figure*}

\subsection{Results}

Fig.~\ref{fig:QKscan}(a) compiles the lowest energy obtained in every \((Q,K_x)\) sector encountered during an \(N_{\mathrm{vac}}\) sweep carried out with a window \(N_{s}=104\) and bond dimension \(\chi=512\). We calibrate \(Q=0\) to the configuration that exactly fills the right half of the variable window and rigidly shift each dispersion so that the ground state sits at \(K_x=0\). The energetically preferred edge carries charge \(Q=-3.5e\); this sector hosts both the ground state and the entire set of low-lying excitations analyzed below.

Curves sharing the same color correspond to different values of $N_{\mathrm{vac}}$.  
In principle, they should collapse onto just two curves, because for a given $(Q,K_x)$ the ground‐state energy depends only on the boundary condition.  
The imperfect collapse in Fig.~\ref{fig:QKscan}(a) originates from the limited window size $N_s$ and bond dimension $\chi$.  
Enlarging the window to $N_s = 360$ and increasing the bond dimension to $\chi = 2048$ in the $Q = -3.5e$ sector restores perfect overlap among the curves for different $N_{\mathrm{vac}}$. 
In both cases, we perform the calculation with the left boundary in both the $0110$ and the $1001$ sector.
These two choices differ by a neutral fermion threading the window and give rise to the pair of dispersion branches shown in Fig.~\ref{fig:QKscan}(b). The lower branch is therefore assigned to the trivial sector $\mathbb 1$, while the higher branch corresponds to the neutral‐fermion sector $\chi$.  
Their splitting at $K_x = 0$ is below $5\times10^{-3}E_{C}$ and decreases further as $\chi$ is increased, consistent with neutral‐fermion modes that remain gapless in the thermodynamic limit.

The ground state supports a charge-density wave with period \(d\approx5\ell_{B}\) and decay length \(L_{e}\approx40\ell_{B}\) (main-text Fig.~2A). To expose its topological structure we evaluate the orbital entanglement spectrum (OES) at \emph{every} bond of the \(\chi=2048\) ground-state MPS; the complete data set is displayed in Fig.~\ref{fig:OESFullgs}. Deep in the bulk the counting is anti-chiral with sequences \((1,1,3,\dots)\) or \((1,2,4,\dots)\), reflecting the two aPf anyon sectors that the cut can select. \emph{At each density maximum} the counting remains anti-chiral aPf counting, whereas \emph{at each density minimum} it becomes chiral--though it does not match the Pf pattern because local hybridization between counter-propagating modes distorts edge modes along entanglement cuts. This alternation persists for three full oscillations before the OES reverts to uniform aPf, confirming a parent configuration of alternating Pf and aPf strips.

Additional insight comes from an excited state in the \(Q=-2e\) sector (main-text Fig.~2B). Its OES, plotted bond-by-bond in Fig.~\ref{fig:OESFullext}(b), shows unambiguous chiral Pf counting at every density minimum, while the maxima retain the anti-chiral aPf structure. Although this state is higher in energy by \(\Delta=0.15E_{C}\), the clean one-to-one correspondence between Pf/aPf counting and density minima/maxima allows us to interpret the \emph{unhybridized parent} to be alternating Pf–aPf strips: the ground state is obtained from the same Pf–aPf strip pattern after partial Majorana-mode hybridization. The stronger and longer-range density oscillation is consistent with the larger excess charge (\(\sim1.5e\)) accommodated at the edge.

\begin{figure}[htbp]
\includegraphics[width=0.4\columnwidth]{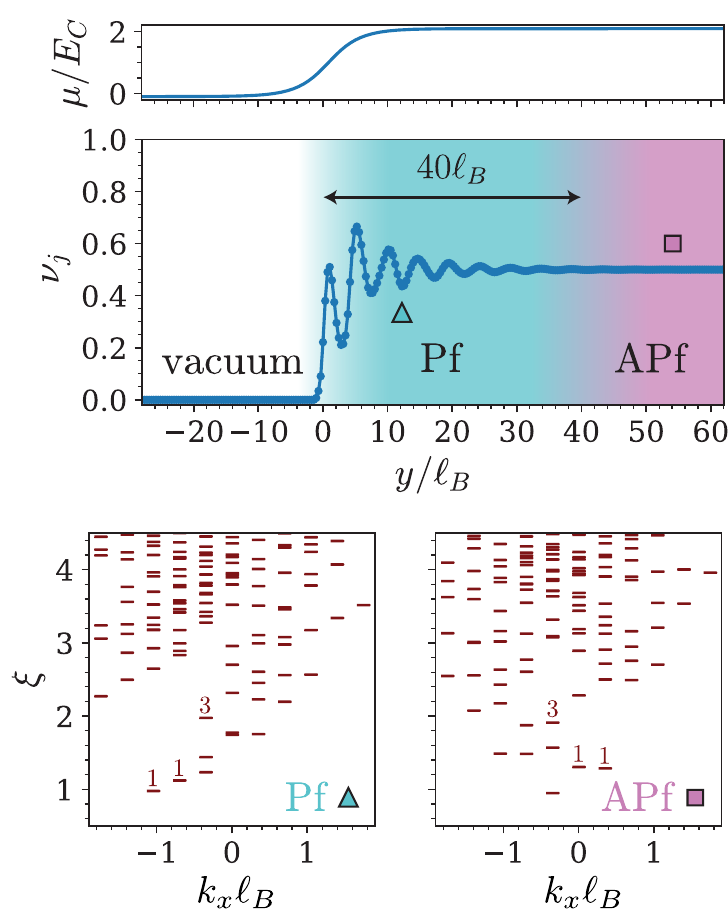}
\caption{\textbf{Edge Reconstruction for an Intermediate Confining Potential.}
(a) Background chemical potential $\mu(y)$ used in the DMRG calculation; the gentle slope corresponds to $\mu_{R}=-0.1E_{C}$ on the vacuum side. 
(b) Orbital filling factor $\nu_j$ obtained for the ground state.  
The shaded region highlights a single Pf strip of width $\sim40\ell_{B}$ that forms between the vacuum and the aPf bulk.  
(c,d) Orbital entanglement spectra (OES) taken at the points marked by the $\triangle$ (Pf strip) and $\square$ (aPf bulk) in panel (b).  
Panel (c) shows uniformly chiral Pf counting $(1,2,4,\dots)$, while panel (d) displays the anti-chiral aPf counting $(1,1,3,\dots)$, confirming complete hybridization of the counter-propagating Majorana modes inside the Pf strip.  
The charge-density oscillation retains its characteristic spatial period $\approx5\ell_{B}$ but with reduced amplitude relative to the sharper-edge case.  
Data were obtained on a cylinder of circumference $L_{x}=18\ell_{B}$ with gate spacing $d=5\ell_{B}$, $\mu_{L}=-0.1E_{C}$, bond dimension $\chi=2048$, and $N_{s}=360$.}
\label{fig:SoftDMRG}
\end{figure}

\section{Intermediate confinement and $\mathcal{C}\neq2$ phases}

To explore an intermediate confining potential, we set the vacuum-side chemical potential to \(\mu_{R}=-0.1E_{C}\) while keeping \(L_{x}=18\ell_{B}\) and \(d=5\ell_{B}\) unchanged.  
The resulting ground state, shown in Fig.~\ref{fig:SoftDMRG}, exhibits a single Pf strip roughly \(40\ell_{B}\) wide at the edge. Across this strip the orbital entanglement spectrum is uniformly chiral with Pf counting \((1,1,3,\dots)\) or \((1,2,4,\dots)\) depending on the entanglement cut, independent of whether the local charge density sits at a maximum or a minimum--evidence that the counter-propagating Majorana modes have hybridized completely.  
The charge-density wave retains its characteristic period \(\approx5\ell_{B}\) but its amplitude is noticeably reduced, consistent with the gentler potential gradient.

A smoothly varying confinement corresponds, in the coupled-stripe picture of Sec.~IV, to stronger tunneling between Majorana modes that are separated by an aPf domain (\(\mathbf g^{1,2},\mathbf g^{3,4},\ldots\)) than between those separated by a Pf domain (\(\mathbf g^{2,3},\mathbf g^{4,5},\ldots\)); cf.\ Eq.~\eqref{eq:gs_def}.  
Departing from the balanced distribution assumed in Eq.~\eqref{eq:randomparametrization} therefore favors the \(\mathcal C=4\) phase over the \(\mathcal C=2\) phase discussed in the main text.  
In the extreme limit where \(\mathbf g^{1,2}\gg\mathbf g^{2,3}\) the model yields a fully hybridized Pf strip, exactly as observed numerically in Fig.~\ref{fig:SoftDMRG}.  
Although the entanglement spectrum no longer resolves individual Pf/aPf domains, the prevalence of chiral Pf counting throughout the oscillatory region is fully consistent with this \(\mathcal C=4\) limit of the stripe construction.

\section{Bloch Hamiltonian and topological invariants}
This section details the calculation of the topological invariant $\mathcal{C}$ within the $H_{4N}+H_c$ part of our model. We start by assuming $N$ is even, in which case the invariant may be interpreted as a Chern number. We enforce a periodic structure with a unit cell of $8$ Majorana modes by requiring $\mathbf{v}^{\alpha}=\mathbf{v}^{\alpha+2}$, $\mathbf{f}^{\alpha}=\mathbf{f}^{\alpha+2}$ and $\mathbf{g}^{\alpha,\alpha+1}=\mathbf{g}^{\alpha+2,\alpha+3}$. Crucially, we allow for $\mathbf{f}^{1}\neq \mathbf{f}^{2},\ldots$ -- there is no symmetry to enforce this. After a Fourier transformation perpendicular to the edge, we get the $8\times8$ Bloch Hamiltonian
\begin{equation}
    H(k,q)= \left[\begin{array}{cc} 
    -\mathbf{v}^1 k + i \mathbf{f}^1 & i \mathbf{g}^{1,2}-i (\mathbf{g}^{2,3})^T e^{-iq} \\
    -i (\mathbf{g}^{1,2})^T+i\mathbf{g}^{2,3} e^{iq}  & \mathbf{v}^2 k + i \mathbf{f}^2
    \end{array}  \right].\label{eq:blochHam}
\end{equation}
While Eq.~\ref{eq:blochHam} obeys $H(k,q)=H(k,q+2\pi)$, there is no such relation involving the momentum $k$ parallel to the edge. Thus the Brillouin zone (BZ) is an infinte cylinder. Noting that $H(k,q)$ becomes trivial at large $|k|$, we can expect a vanishing Berry curvature there, so we may still define a Chern number for this BZ. The Hamiltonian $H(k,q)$ has 8 bands and is particle-hole symmetric -- we compute the sum of the Chern numbers of the 4 negative-energy bands, for which we use the non-Abelian version of the algorithm from \cite{78_Fukui_2005}. We use the distribution in Eq.~\ref{eq:randomparametrization} of the main text to sample $\mathbf{f}^1,\mathbf{f}^2,\mathbf{g}^{1,2}$ and $\mathbf{g}^{2,3}$, while we choose $\mathbf{v}^1=\mathbf{v}^2=\mathbb{1}_4$ throughout as we observe no qualitative change in the distribution when speeds are allowed to vary.

\subsection{Adding top modes} \label{sect:top_modes_to_quads}
The Chern number argument above considers only a subset of the modes at the interface. The others must be added by hand, which we always do assuming a strong local coupling. Near the edge, there is the additional mode $\psi_0$ -- when combined with the modes originating from the oscillating region via $H_{c0}$, we find that there are $\mathcal{C}-1$ upstream Majorana modes near the edge. Further in the bulk, we find $\mathcal{C}$ downstream modes if $N$ is even and $4-\mathcal{C}$ upstream modes if $N$ is odd. In the latter case, the $N$-th quadruplet interacts with the $\mathcal{C}$ chiral modes via $\mathbf{g}^{N-1,N}$. Note that in all cases, the total central charge is independent of $\mathcal{C}$ and matches the value mandated by the physical edge.

\begin{figure}
     \centering
     \begin{subfigure}[b]{0.24\textwidth}
         \centering
         \includegraphics[height=1.4in]{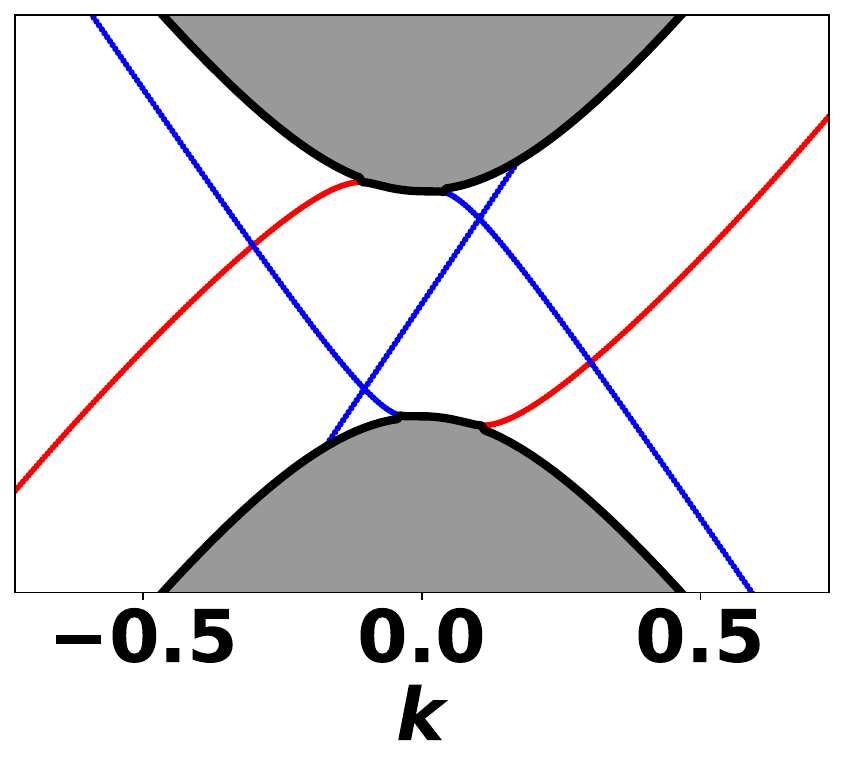}
         \label{fig:q1_top_0}
     \end{subfigure}
     \hfill
      \begin{subfigure}[b]{0.24\textwidth}
     \centering
     \includegraphics[height=1.4in]{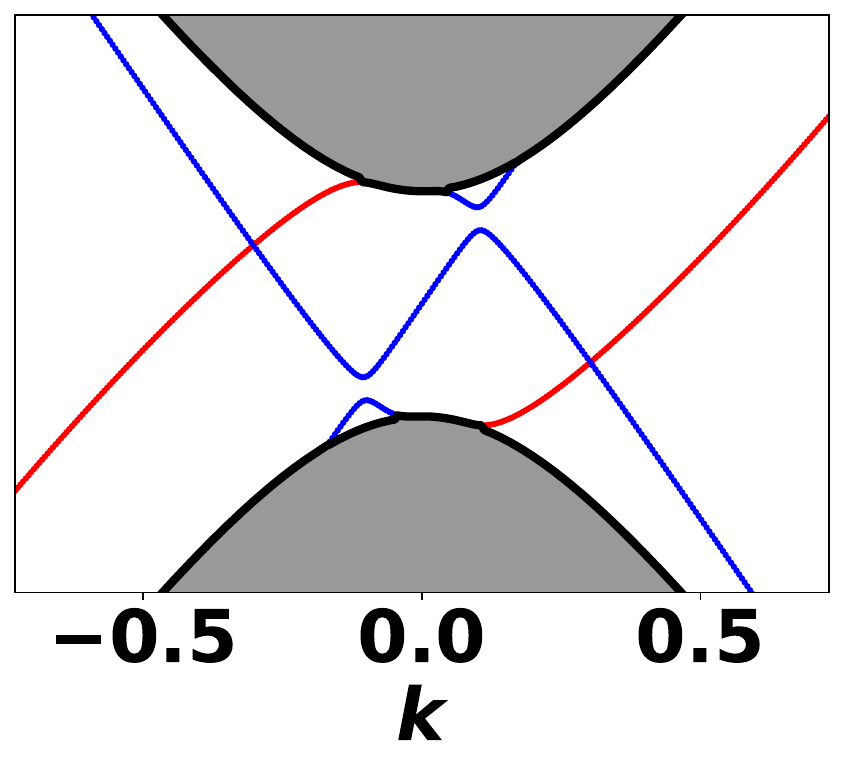}
     \label{fig:q1_top_1}
     \end{subfigure}
     \hfill
      \begin{subfigure}[b]{0.24\textwidth}
     \centering
     \includegraphics[height=1.4in]{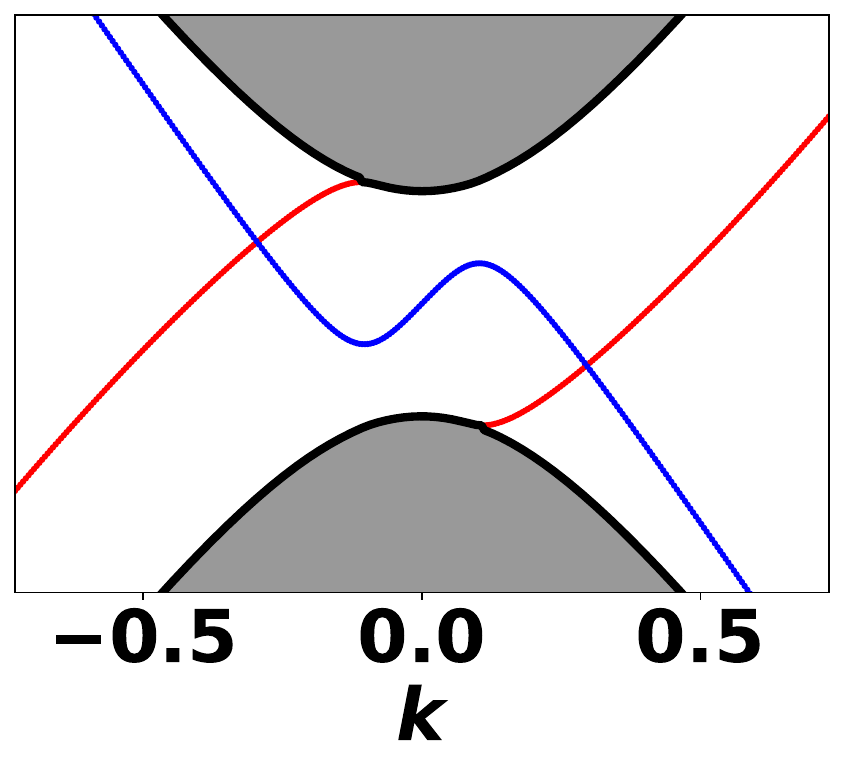}
     \label{fig:q1_top_2}
     \end{subfigure}
     \hfill
      \begin{subfigure}[b]{0.24\textwidth}
     \centering
     \includegraphics[height=1.4in]{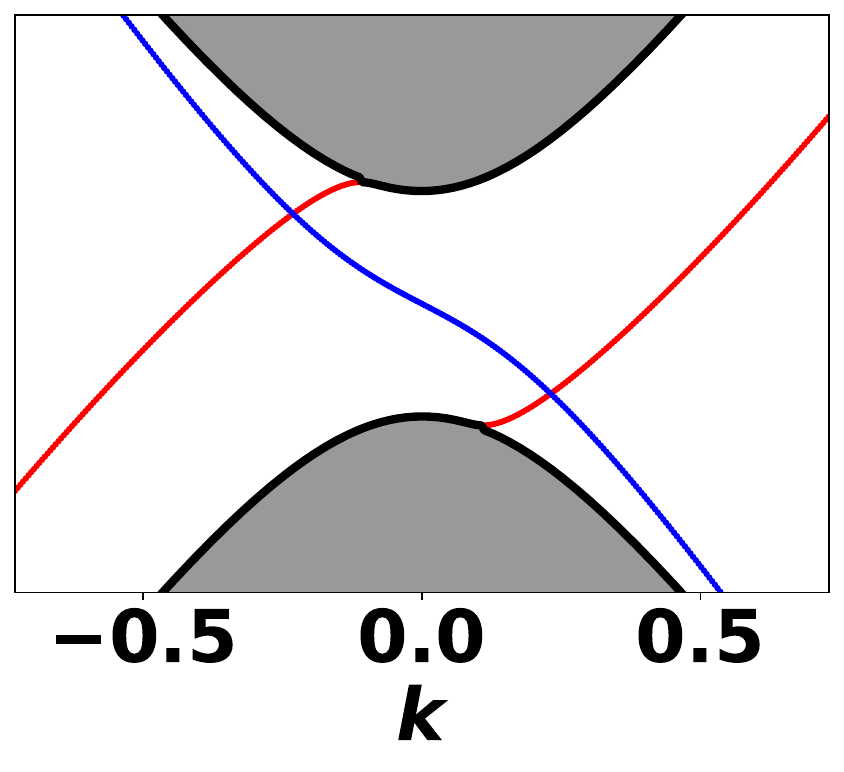}
     \label{fig:q1_top_3}
     \end{subfigure}
     \hfill
     \vspace{-5px}
     \caption{Adding the top mode to an even-$N$ state with $\mathcal{C}=2$ to model a Pf-2 (aPf-3) interface. Blue modes are at the system edge while red modes are deeper in the bulk. In the leftmost frame, the top mode (positive-sloped blue line) is not interacting with the rest. From left to right, the interaction strength $H_{c0}$ of the top mode with the rest is increased and the modes gap to produce one upstream edge Majorana mode. Without edge reconstruction, we would only see one downstream mode at the system edge (one positive-sloping blue line). The Pf-3 (aPf-2) interfaces have the deep (red) mode directions reversed.}
     \label{fig:q1_adding_top_modes}
\end{figure}

To match experimental results without fine-tuning, we require that $c_\text{Edge}$, the central charge of all the modes near the edge (including the charged boson, but not including the deep modes) obeys $\left| c_\text{Edge}\right|=\frac12$ for interfaces with both $\nu=2$ and $\nu=3$ IQHE. It may be checked that this only holds for the $\mathcal{C}=2$ state. In Fig.~\ref{fig:q1_adding_top_modes} we take a representative $\mathcal{C}=2$ state with $N$ even and add the top chiral Majorana mode. As the interaction strength $H_{c0}$ is increased, we see how the top mode gaps with the two edge modes of $H_{4N}$. Crucially, this has no effect on the two chiral modes deeper in the bulk. Finally, the single remaining Majorana mode near the edge must have momentum $k=0$ while the two deeper modes have $k=\pm k_0$ -- this momentum mismatch is part of the reason why these modes struggle to equilibrate.

\section{Transfer-matrix Analysis for Quenched Disorder}
The above discussion has found that the Majorana mode system may likely be in a phase with $\mathcal{C}=2$, which, when combined with the additional Majorana mode on the FQHE-vacuum interface results in an interface compatible with experimental observations. But this has relied on two assumptions -- that all oscillations are the same (i.e. translational invariance perpendicular to the edge) and that the system is translationally invariant parallel to the edge. Although the Majorana interactions we discussed are not solely due to quenched disorder, its inclusion could significantly affect the outcome of our calculations. The topological nature of the result, however, suggests some degree of stability against perturbations. In this section, we simulate systems where both forms of translational invariance are explicitly broken by quenched randomness. To do this, we take a finite system and use a transfer-matrix method analogous to the methods used in \cite{23_chalker_PhysRevB.65.054425}. For a system of $M$ Majorana modes $\psi_1,\ldots,\psi_M$, the Lagrangian in real space may be written as (summation understood)
\begin{equation}
    \mathcal{L}=i \psi_a \left(\partial_t + v_a \delta_{ab} \partial_x - M_{ab}(x)\right)\psi_b
\end{equation}
Here, $v_a$ are the signed velocities of the modes, while $M_{ab}(x)$ is the real skew-symmetric interaction matrix which is now allowed to depend on position $x$. Instead of a smooth variation of $M(x)$, we discretize the space into intervals separated by points $\{x_n\}$. For $x_{n-1}<x\leq x_n$, take $M(x)=M_n$. Then, we have the full system Lagrangian 
\begin{equation}
    L = \int\text{d}x \mathcal{L}(x) = \sum_n \int_{x_{n-1}}^{x_n}\text{d}x \Psi^T\left(i\partial_t + iV\partial_x - iM_n\right)\Psi
\end{equation}
where $\Psi=(\psi_1,\ldots,\psi_M)$ and $V_{ab}=\delta_{ab}v_a$ is the diagonal matrix of the signed velocities. We are interested in the low-energy physics, so we may set $\partial_t\Psi=0$ in the Lagrangian. Then the mean field is simply read off $\partial_x\Psi = (V^{-1})M_n\Psi$. We assume all velocities to be nonzero, so $V$ is invertible. The evolution along the $x$-axis is in general not unitary if there are counter-propagating modes -- we may solve for the evolution to give
\begin{equation}
\begin{split}
    \Psi(x_K)=\exp\left(\left[x_K-x_{K-1}\right]V^{-1}M_K\right)\exp\left(\left[x_{K-1}-x_{K-2}\right]V^{-1}M_{K-1}\right) \ldots \\ \ldots
    \exp\left(\left[x_1-x_0\right]V^{-1}M_1\right)\Psi(x_0)=T_K\Psi(x_0).
\end{split} \label{eq:transfer_matrices_multiplied}
\end{equation}
This defines a transfer matrix $T_K$. We are physically interested in the limit of $K\to \infty$, where we keep the difference of consecutive $x_n$ at order a few $\ell_B$, or shorter. A mode localised over a lengthscale $\xi$ shows up as an eigenvalue of magnitude $|\lambda| \sim e^{\pm (x_K-x_0)/\xi}$ in $T_K$. So the eigenvalue associated with a localised state may be either exponentially large or exponentially small in the distance over which the transfer matrix is computed. Delocalised modes should have eigenvalues with $|\lambda|=1$.

To make the large-$K$ limit numerically accessible, we use the Lyapunov-exponent-based algorithm of \cite{23_chalker_PhysRevB.65.054425}. Let $T_n=\exp\left(\left[x_n-x_{n-1}\right]V^{-1}M_n\right)$ be the transfer matrix associated with the $n$-th interval. If the elements of $T_n$ are not too large, we can multiply multiple consecutive transfer matrices to get $\mathcal{T}_{(n+r,n)}=T_{n+r-1}T_{n+r-2}\ldots T_n$. The number $r$ of matrices multiplied should be as large as possible while keeping a specified numerical precision. The main idea is to use $\mathcal{T}_{(n+r,n)}$ to act on a set of eigenvector guesses, noting the growth rates $\lambda_k$ (Lypunov exponents $\epsilon_k=\log \left|\lambda_k\right|$). After that, Gram–Schmidt orthonormalization is performed and the process repeated until we see convergence of the Lypunov exponents. An important quantity is the localisation length of a mode, defined by $\ell_k = r \Delta x / \langle \epsilon_k\rangle$, where we assumed that the intervals are evenly spaced, $x_n-x_{n-1}=\Delta x~\forall n$.

We take the matrix structure as in Eq.~\ref{eq:bigmatrixH} and take the elements, which are now functions of position in a Gaussian distribution analogous to Eq.~\ref{eq:randomparametrization}. But we allow for a finite correlation length between the matrix elements with an exponential decay, 
\begin{equation}
    \begin{split}
        \left\langle g_{l_1l_2}^{\alpha_1,\alpha_1+1}(x)g_{l_3l_4}^{\alpha_2,\alpha_2+1}(x')\right\rangle&=\delta_{l_1l_3}\delta_{l_2l_4} R^{|l_1-l_2+s-3|} e^{-|x-x'|/\xi_1} e^{-|\alpha_1-\alpha_2|/\xi_2}\\
        \left\langle f_{l_1l_2}^{\alpha_1}(x)f_{l_3l_4}^{\alpha_2}(x')\right\rangle&=\delta_{l_1l_3}\delta_{l_2l_4} R^{|l_1-l_2|-1} e^{-|x-x'|/\xi_1} e^{-|\alpha_1-\alpha_2|/\xi_2}\,\,\,\,(l_1<l_2,~l_3<l_4)
    \end{split}
\end{equation}
where $\alpha_1-\alpha_2 = 2n$ for $n\in\mathbb{Z}$. If $\alpha_1-\alpha_2$ is odd, there is no correlation. Setting $\xi_1,\xi_2\rightarrow\infty$ would recover the method of using topological invariants. The choice of an exponential decay of correlations in position space is one of many possibilities, which is chosen for its numerical convenience. Furthermore, we choose all the velocities to be of the same magnitude -- we have observed that allowing them to vary has no significant effect on the results. Setting their magnitude to unity implies a choice of the unit of length $\sim \ell_B$.

We choose to focus on a finite system with 13 (14) Majorana mode quadruplets in order to model the aPf-2 (Pf-2) interfaces. We choose a correlation length of disorder $\xi_1= 20$ and we assume $\xi_2 \sim 3$ -- this is motivated by expecting $\xi_1$ and $\xi_2$ to be at the same scale (as the period of the oscillations is $\sim 6 \ell_B$).

\begin{figure}
     \centering
     \begin{subfigure}[b]{0.33\textwidth}
         \centering
         \includegraphics[height=2.3in]{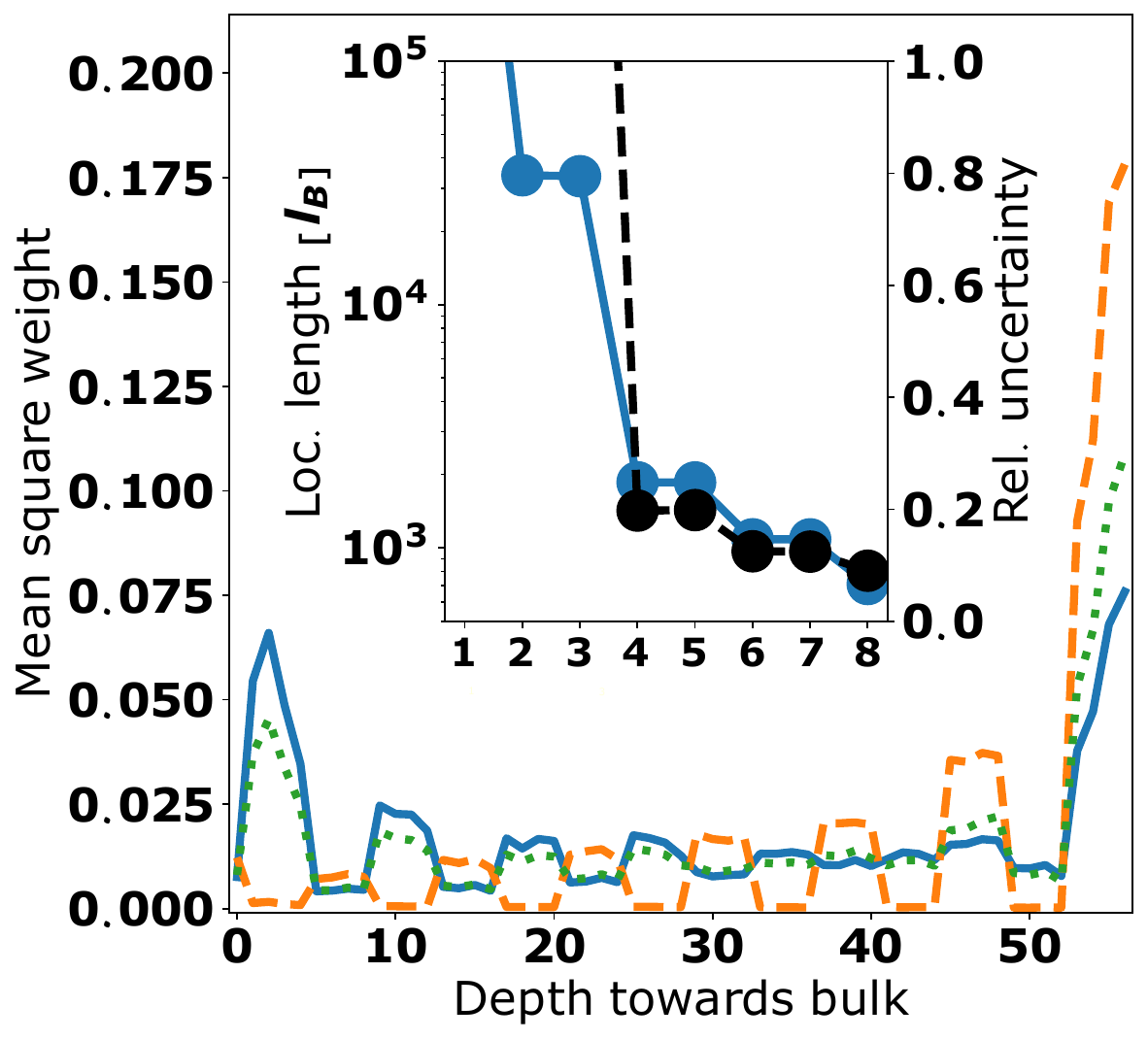}
         \caption{The Pf-2 interface with 14 quadruplets of Majorana modes}
         \label{fig:pf2_transfermatrix}
     \end{subfigure}
     \hspace{1in}
     \begin{subfigure}[b]{0.33\textwidth}
         \centering
         \includegraphics[height=2.3in]{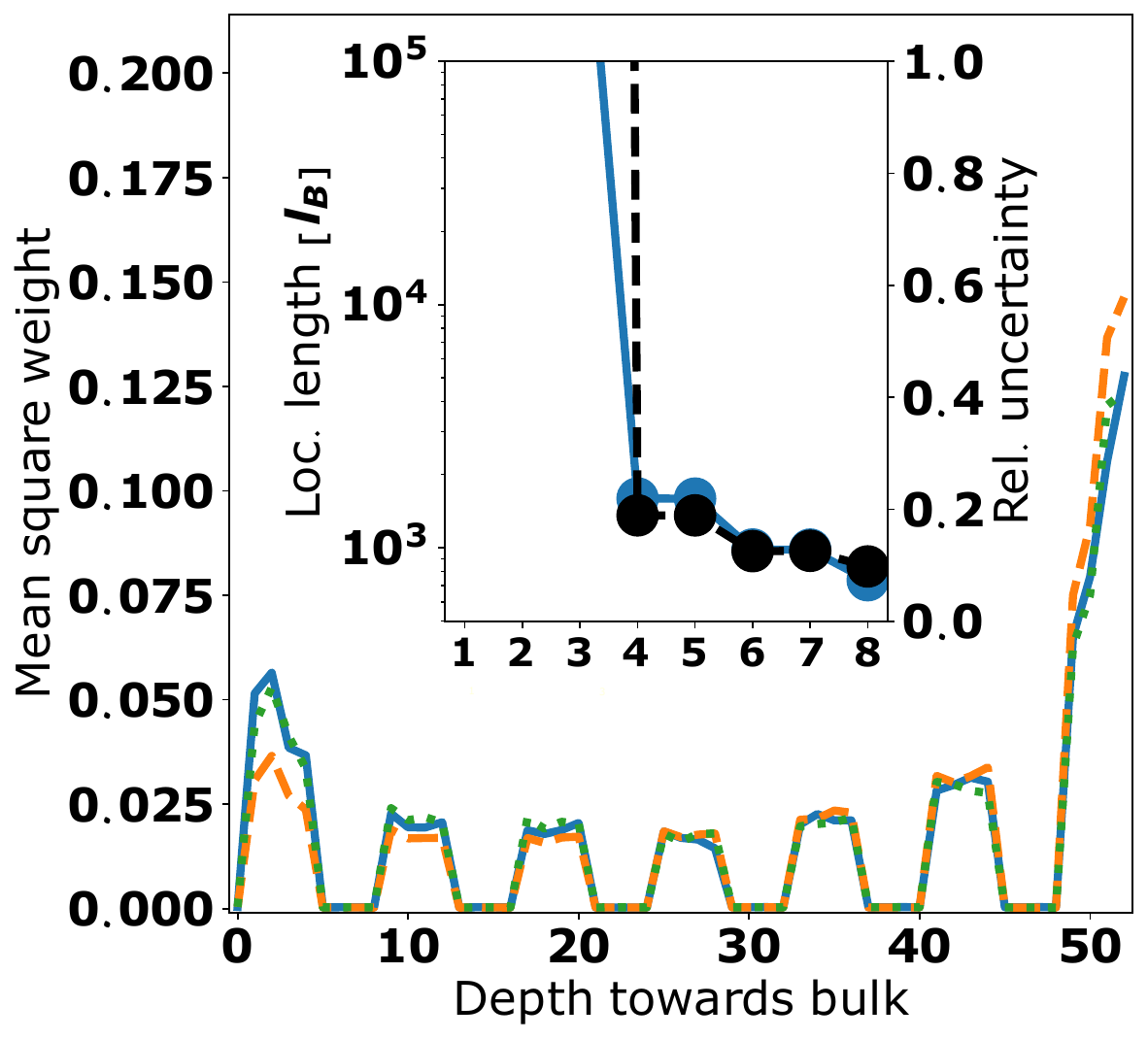}
         \caption{The aPf-2 interface with 13 quadruplets of Majorana modes}
         \label{fig:aPf2_transfermatrix}
     \end{subfigure}
     \caption{The square weights of the three delocalized weights on both interface types. Insets show localization lengths (solid blue) and their relative uncertainties (dashed black) for the different modes. In both cases, the first three modes have a very large relative uncertainty and with a localization length larger than others by an order of magnitude. The weights of these three eigenmodes of very large localization length are shown on the main axes.
     }
     \label{fig:transfer_matrix_results}
\end{figure}

With this, we run the above simulation over a distance of $5\cdot10^5$ $\ell_B$ to find the eigenvectors and localization lengths of various modes. When $s=0$ and $R<0.6$ or when $s=1$ and $R<0.4$, a particular structure emerges in the results -- for both types of interface, three modes seem to be delocalised, while the rest are well-localized. Fig.~\ref{fig:transfer_matrix_results} shows the example of $R=0.2$ and $s=1$ for both interface types -- the eigenvector weights show a picture where there is some weight close to the edge and approximately double the weight near the bulk side. The localization lengths and their relative uncertainties are shown in the insets -- for three modes in each case, the uncertainties are larger than unity, indicating a high likelihood that these modes are truly delocalized. These modes strongly stand out from the rest. Similar results are found for the values of $s$ and $R$ described above, as well as for different correlation lengths.

The eigenvectors found here are a linear combination of the three modes discussed -- but note that on short length scales, they are in fact degenerate (have the same Lypunov exponent of zero) -- so we may take the modes as any desired linear combination. But on scales comparable to $l_\text{eq}$ of these modes, these combinations must be considered. This indicates that the modes mix over distances similar to $l_\text{eq}$ which are $\gtrsim 10^4 \ell_B\sim 1000~\mu\text{m}$, which is significantly longer than the lengths in experiments (tens of $\mu$m). Other parameter ranges, such as $s=2$ or having $R\sim 1$ do not show such results. But it should be noted that this assumes the parameters are entirely spatially random and of zero mean -- in reality, they may have a large constant component with random noise superimposed -- in that case, the system may even more reliably display this behaviour, provided that the constant component puts it into the $\mathcal{C}=2$ phase. But even a fully random system can generically display the type of behaviour discussed in the main text -- an upstream mode near the system edge and two Majorana modes moved deeper towards the bulk.

\subsection{Random operators and their scaling properties} \label{sect:more_on_rg}
Say that the Lagrangian includes a term $\xi(x) \mathcal{O}(x,t)$ where $\mathcal{O}$ is an operator of scaling dimension $\Delta$ and $\xi(x)$ is spatially random noise with ensemble average $\left\langle \xi(x)\xi(x')\right\rangle_\text{dis} = W\delta(x-x')$. The scaling dimension $\Delta$ is defined by requiring $\left\langle \mathcal{O}(x,0)\mathcal{O}{x,t}\right\rangle \sim t^{-2\Delta}$ for $t$ large. Then, to the lowest order, the RG-flow equation is \cite{07_PhysRevLett.125.016801,11_PhysRevB.51.13449,cite_for_random_rg_eqs_PhysRevB.99.085309}
\begin{equation} \label{main_eq:scaling_rg}
    \frac{\text{d}W}{\text{d}l}=\left(3-2\Delta\right)W,
\end{equation}
showing that operators with $\Delta \leq \frac32$ are relevant. Such spatially random operators are used to include disorder in our calculations and they are generally what leads to thermal equilibration of the modes. We may follow the reasoning of \cite{07_PhysRevLett.125.016801}, where Eq.~\ref{main_eq:scaling_rg} is used to show that an operator with scaling dimension $\Delta$ and a random pre-factor will lead to equilibration of the edge modes over a distance $l_\text{eq}\sim l_c \left(l_T / l_c \right)^{2\Delta-2}$ -- since $l_T$ diverges at low temperatures, operators with $\Delta>1$ will equilibrate weakly as the temperature is lowered. For low enough temperature, $l_\text{eq}$ surpasses the experimental system size and then the modes fail to equilibrate. This growth in $l_\text{eq}$ and consecutive transition into non-equilibrium is more pronounced when $\Delta$ is larger. For non-random operators the scaling is different, $l_\text{eq}\sim T^{3-2\Delta}$. This is generally shorter than for the same operator with a random prefactor, so we might expect it to dominate. But a point to consider is that nonrandom operators have to conserve energy and momentum, which is not possible for an operator connecting two modes with different velocities \cite{07_PhysRevLett.125.016801}. This significantly suppresses the contribution of nonrandom operators to equilibration. Of course, a microscopic value of $W$ must still be input to the theory. Assuming that independent scatterers are a distance $l_\text{Disorder}$ apart and that each has dimensionless amplitude $S$ (the  pre-factor in front of the scattering term beyond the necessary dimensionful factors, expected to be $\mathcal{O}(1)$), it can be shown that
\begin{equation} \label{eq:best_eq_lenght_general_eq}
    l_\text{eq, random} \sim l_\text{Disorder} \frac{1}{S^2} \left(\frac{l_T}{l_c}\right)^{2\Delta-2},
\end{equation}
with $l_T=\frac{v}{T}$ the thermal length. This form  proves to be the most convenient in what follows.

\section{Approximate experimental parameters} \label{sub:experiment_values}
While the different experiments \cite{48_Banerjee_2018,50_Dutta_2022,25_paul2023topological,with_noise_experiment} differ in size and exact details, all generally have contacts of size $\sim 10~\mu\text{m}$, which is also roughly the scale of separation between the contacts over which the edge modes propagate. The experiments are done around $B\sim 5\text{ T}$ which puts the magnetic length at $\ell_B\sim 12\text{ nm}$. The experiments are done around $T\sim 15\text{ mK}$. While the exact velocities of the edge modes are not known, taking the Majorana mode velocity $v_\psi\sim 10^6 \text{ cm}/\text{s}$ and charged mode velocity $v_\phi\sim 10^7 \text{ cm}/\text{s}$ seems reasonable \cite{03_PhysRevLett.124.126801,02_PhysRevB.99.085309,25_paul2023topological,a_majorana_mode_velocity_thing_PhysRevB.80.235330}. This puts the thermal length at $l_T=\frac{\hbar v_\psi}{k_B T}\sim 5~\mu\text{m}$ for the Majorana modes, and at $\sim 50~\mu\text{m}$ for the charged modes. The length scale associated with pairing in the states at $\nu=\frac52$ is $l_c\sim 8\ell_B\sim 100\text{ nm}$ \cite{12_PhysRevLett.125.146802}. There is always some disorder present as the semiconductor (for example GaAs) in which the experiment is performed must be doped to be able to put electrons in the 2D quantum well in which the FQHE state lives. From the depth profile shown in \cite{50_Dutta_2022}, we may estimate that most of the disorder is situated $\sim 100\text{ nm}$ above/below the state so we may take this to be the disorder correlation length \cite{12_PhysRevLett.125.146802}. Doping densities range around $10^{11} \text{ cm}^{-2}$, which corresponds to a mean spacing $\sim 30\text{ nm}$ between dopants.

\section{Decoupling of the inner Majorana modes} \label{sect:mm_tunneling}
Consider two Majorana modes, where tunneling between them is allowed. When the operator $\mathcal{O}$ is taken as Majorana mode tunneling $\mathcal{O}=S \psi_a \psi_b$, we can see that $l_\text{eq}\sim l_\text{Disorder}/S^2$ due to the scaling dimension $\Delta_\mathcal{O}=1$. When propagating over lengths much shorter than $l_\text{eq}$, the modes are effectively thermally decoupled. They are thermally coupled over lengths larger than this scale. While a short-ranged impurity may be represented by a term $S \Gamma l_c \delta(x) \mathcal{O}(x)$ in the Lagrangian (here $S$ is taken as dimensionless and $\Gamma l_c$ is the factor required on dimensional grounds), it turns out to be beneficial to also consider impurities extended over a distance $w_D$ which may be represented by a term $\Gamma f(x) \mathcal{O}(x)$ where $f(x=0)\sim 1$ and $f(x)\rightarrow0$ as $|x|/w_D\gtrsim 1$. It turns out that the argument above continues to hold provided that $w_D\ll l_T$, where we simply have to make the replacement
\begin{equation} \label{eq:transfer_to_wide}
    S\rightarrow  \left|\frac{1}{l_c}\int \text{d}x f(x)\right|.
\end{equation}
As discussed above, experiments generally have $l_T\sim 5~\mu\text{m}$, while disorder is only correlated on scales $\sim 100\text{ nm}$.

With these pre-requisites, we are ready to tackle the problem of the three Majorana modes which survive in our system after the rest have been gapped out ny the inter-wire couplings. We will assume that the system is in the $\mathcal{C}=2$ phase. In that case, the edge is effectively described by a downstream $\nu=\frac{1}{2}$ chiral boson $\phi$, an upstream Majorana mode $\chi$ localised at the edge and two inner Majorana modes $\psi_{1,2}$ propagating upstream (downstream) at aPf-2/Pf-3 (aPf-3/Pf-2). These may be combined into the Lagrangian
\begin{equation} \label{main_eq:reconstructed_edge_lagrangian}
        \mathcal{L} = \frac{2}{4\pi}\partial_x\phi(-i\partial_\tau + v_\phi\partial_x)\phi+\chi(\partial_\tau - i v_\chi \partial_x)\chi+ \Psi^T (\partial_\tau - i s v_\psi \partial_x + v_\psi k_0 \sigma_y)\Psi.
\end{equation}
Where $s=1$ ($s=-1$) for aPf-2 (Pf-2) and $\Psi=(\psi_1,\psi_2)^T$ is a vector containing the two inner Majorana modes. Note the term $\Psi^T v_\Psi k_0 \sigma_y \Psi=-2ik_0\psi_1\psi_2$ -- as discussed, this gives a finite momentum $k=\pm k_0 \sim \ell_B^{-1}$ to the modes.

To show that this edge model can explain the observed edge thermal conductance, we first argue that due to a combination of the lack of wavefunction overlap and momentum mismatch the inner modes entirely decouple from the thermal transport in experiment -- not only do they not equilibrate with the other modes, but they do not equilibrate with the thermal contacts either. 

The most relevant operator that could lead to equilibration between the inner modes and the edge is Majorana mode tunnelling $\mathcal{O}=\chi\psi_{1,2}$. It has scaling dimension $\Delta=1$, which according to Eq.~\ref{main_eq:scaling_rg} makes it RG-relevant. Consider the general tunneling perturbation
\begin{equation} \label{main_eq:inner_modes_perturb_before_gauge_trans}
    \mathcal{L}_\text{tunnel}=i\frac{\sqrt{v_\psi v_\chi}}{l_c}G\chi F^T(x) \Psi
\end{equation}
to Eq.~\ref{main_eq:reconstructed_edge_lagrangian}. Here, $F(x)=\left(f_1(x),f_2(x)\right)^T$ represents the coupling of both the inner modes to the edge mode with $f_{1,2}$ smooth functions varying over a distance $\sim w_D$. Here, $G$ indicates the reduction in the tunnelling amplitude between the two modes due to the lack of wavefunction overlap -- this is the factor that depends on the number of quadruplets, $N$. Given that we know the edge modes are localized, we may expect $G\sim e^{-N/\xi}$ for some localization length $\xi$, but this implicitly assumes disorder only allows nearby modes to interact -- there might be some longer-range disorder, leading to a less sharp reduction in $G$ as a function of $N$. We must first remove the non-random coupling between $\psi_{1,2}$ in Eq.~\ref{main_eq:reconstructed_edge_lagrangian} by an appropriate wavefunction transformation. Letting $\Psi = R(x)\tilde{\Psi}$ with $R(x)\in SO(2)$, such that $is\partial_x R(x) = k_0\sigma_yR(x)$, so that $R(x)=\exp(is\sigma_y k_0 x)\in SO(2)$, the final term in Eq.~\ref{main_eq:inner_modes_perturb_before_gauge_trans} becomes $\tilde{\Psi}\left(\partial_\tau - isv_\psi\partial_x\right)\tilde{\Psi}$, so the interaction term has been removed. In Eq.~\ref{main_eq:inner_modes_perturb_before_gauge_trans}, $F(x)$ is replaced by
\begin{equation} \label{main_eq:gauge_transformed_inner_perturb}
    \tilde{F}(x)=\begin{pmatrix}
        \tilde{f}_1(x)\\
        \tilde{f}_2(x)    
    \end{pmatrix}=
    \begin{pmatrix}
        \cos(k_0x)f_1(x)-\sin(k_0x)s f_2(x) \\ 
        \cos(k_0x)f_2(x)+\sin(k_0x)s f_1(x)
    \end{pmatrix}.
\end{equation}
These transformed forms of the functions $\tilde{f}_{1,2}$ can then be used in Eq.~\ref{eq:transfer_to_wide}. While exact details depend on the form chosen for $f_{1,2}$, we expect the general conclusions to hold for all slowly varying functions $f$ peaked around $x=0$ and with width $w_D$. We choose to parametrise the disorder by a Lorenzian, letting $f_1(x)=f_2(x)=S\frac{w_D^2}{x^2+w_D^2}$. This gives
\begin{equation} \label{main_eq:edge_inner_final}
    l_\text{Edge-inner} \sim  \frac{1}{G^2 S^2} l_{\text{Disorder}} \frac{l_c^2}{w_D^2} e^{2w_D k_0}.
\end{equation}
It is evident that if $w_Dk_0$ is large, the correlation length becomes too big to equilibrate in practice. A Gaussian $f_{1,2}$ would lead to $\sim e^{(w_Dk_0)^2}$ which has even more suppression. While the value of $k_0$ is not known, we expect it to be $k_0 \sim \ell_B^{-1}$. So equilibration due to disorder with large $w_D$ is exponentially suppressed in $w_D/\ell_B$ and will thus be ineffective. If $k_0=0.7 \ell_B^{-1}$ and if we take $l_\text{Disorder}=w_D=100\text{ nm} \sim 8\ell_B$ for these lengths and if we take $S=1$ and $G=1$ (the second being significantly too large), we still see $l_\text{Edge-inner}\sim 3\cdot 10^3~\mu\text{m}$, which is over $10^2$ times larger than typical experiments. But this has not yet taken into account the suppression due to the lack of wavefunction overlap -- in the case where disorder introduces long-range interactions, $G$ is not necessarily exponentially small in the number of quadruplets $N$, but it may obey some power law -- for example the Coulomb-like $G\sim 1/N$. Consider this longer-range power law first. In the Coulomb case, another factor of $N^2 \gtrsim 10^2$ is introduced into the equilibration length, pushing it to $10^4$ times larger than typical experiments -- now even if the impurity strength $S$ is larger, it is unlikely that such disorder can equilibrate the system. In fact we may estimate the strength of $S$ as follows: the disorder is due to charged donors which are $h_\text{dis}\sim 100\text{ nm}$ away from the sample and the typical spacing between them is also $\sim 50\text{ nm}$ \cite{50_Dutta_2022}. We may thus expect that each point feels the potential of roughly one donor. The presence of these donors then alters the local chemical potential via the Coulomb interaction by $\Delta\mu \sim \frac{e^2}{4\pi\epsilon\epsilon_0 h_\text{dis}}\sim 10~v_\psi/l_c$, using $\epsilon\sim 13$ as is appropriate for GaAs \cite{a_majorana_mode_velocity_thing_PhysRevB.80.235330}. If the modes were charged, this would suggest that $S\sim 10$ should be taken in Eq.~\ref{main_eq:edge_inner_final}, but since the modes are uncharged, they may interact more weakly with these fluctuations in the background electrical potential, potentially justifying a smaller $S$. Even with $S\sim 10$ and the other parameters as above, Eq.~\ref{main_eq:edge_inner_final} suggests $l_\text{Edge-inner}\sim 8\cdot 10^2~\mu\text{m}$, which as we see later can still be long enough for the modes to not equilibrate in practice. For this Coulomb-based effect, using $f(x)=S\frac{w_D}{\sqrt{x^2+w_D^2}}$ might be more appropriate. This form leads to an equilibration length $\sim3$ times larger than what is suggested by Eq.~\ref{main_eq:edge_inner_final} for the Lorentzian. This leads us to believe that short-ranged disorder might be more effective at equilibration. In that case however we may expect $G$ to be exponentially suppressed in $N$, ie $G\sim e^{-N/\xi}$ (ie proportional to the wavefunction overlap), again leading to a large equilibration length. Furthermore, disorder with correlation lengths below $w_D\sim 100\text{ nm}$ is not expected anyway, since the layer of doping responsible for the disorder is vertically separated from the electron gas in experiments

This all shows we have good reason to believe the inner modes do not equilibrate well with the other edge modes -- but such non-equilibration generally increases the thermal conductance of an edge \cite{01_PhysRevB.107.245301}. In order for the inner modes to fully decouple, they should also decouple from the floating contacts used to measure the conductance. We expect this to happen for reasons similar to the above. First, since the modes are deeper into the sample, they may not interact with the contacts as strongly. Second, we can again expect that the term coupling the contacts to the edge is smooth (since contacts are large compared to microscopic length scales). Taking the Fourier component at $\pm k_0$ will again reduce the amplitude. We now address the question of how we couple to the contact in more detail. We expect that the contacts couple via electon tunneling. An electron operator at the FQHE edge is built from the Bose mode $\phi$ and a Majorana mode $\psi$, giving $\psi_\text{el}=\psi e^{2i\phi}$. Within our model, any interface contains three Majorana modes and any of these three could take the role of $\psi$, but as discussed above, the two deep Majorana modes have a vanishing wavefunction overlap with the edge Bosonic mode -- this leads us to conclude that any process that includes tunneling an electron into the charged mode and a deep Majorana should be significantly surpressed. This leads us to assume that the two modes at the edge are coupled well to the contacts, while the two inner Majorana modes are not. The exact details of how much weaker the inner modes couple depend on the microscopic physics of the contacts. Crucially any term in the Lagrangian that couples to the contacts suffers from the same two bare suppresion effects as the equilibration terms discussed above, which has led us to the estimate that the lengthscale associated with this coupling is at the same order of magnitude as the equilibration length between the deep modes and the edge.

\subsection{Hydrodynamic edge simulations}
\begin{figure}
     \centering
     \begin{subfigure}[b]{0.48\textwidth}
         \centering
         \includegraphics[width=2.7in]{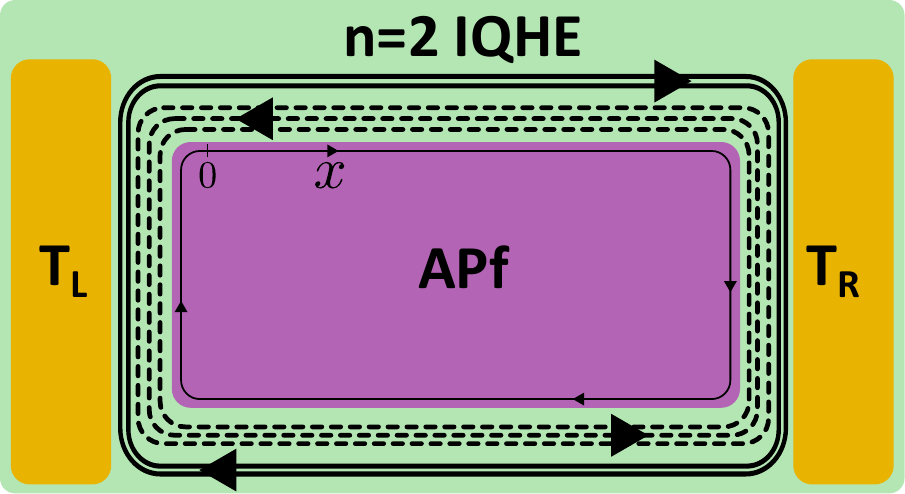}
         \caption{}
         \label{main_fig:hydrodynamic_setup}
     \end{subfigure}
     \hfill
     \begin{subfigure}[b]{0.48\textwidth}
         \centering
         \includegraphics[width=2.7in]{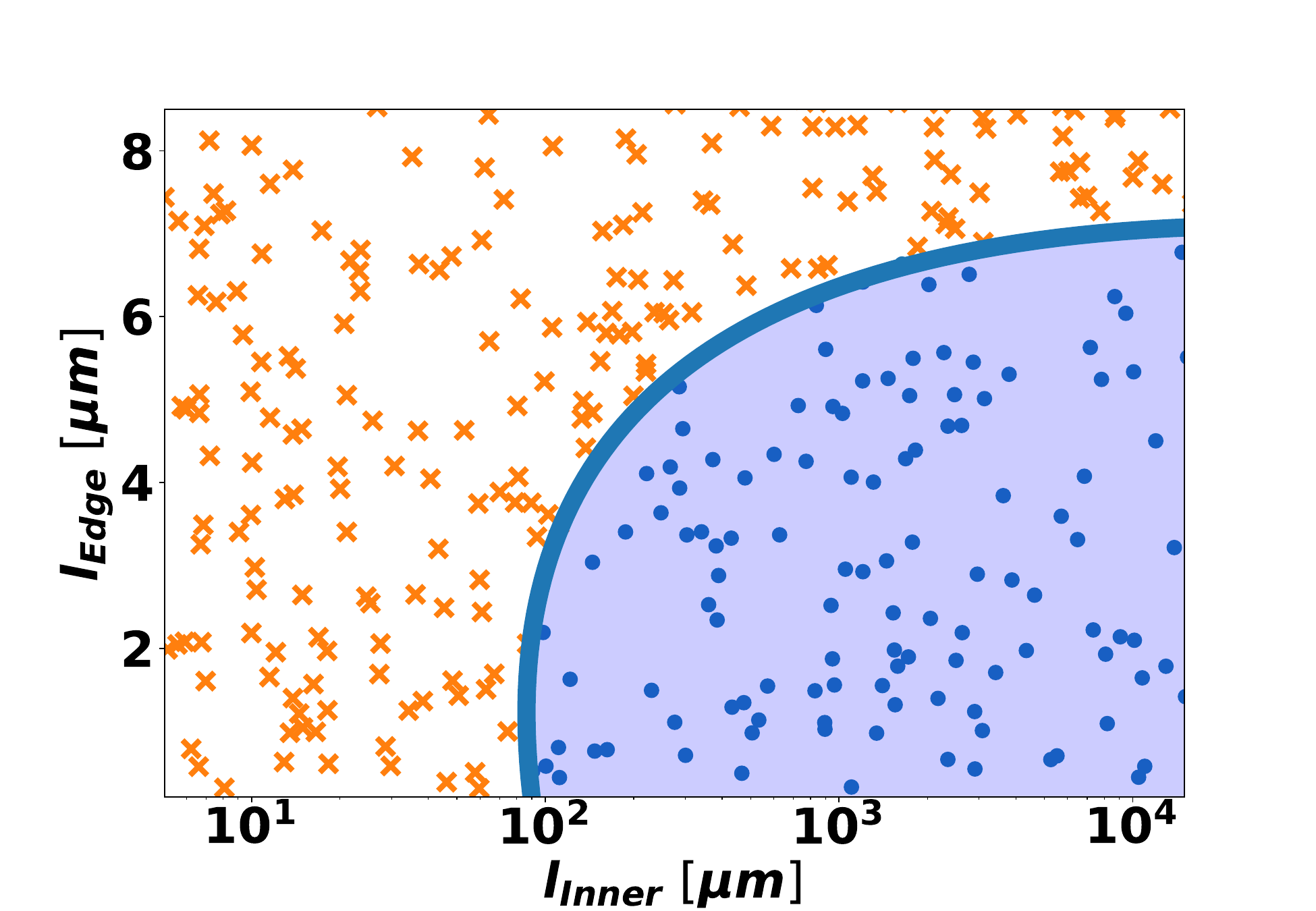}
        \caption{}
        \label{main_fig:range_of_validity}
     \end{subfigure}
     \caption{(a) The setup used in hydrodynamic simulation for the aPf-2 case. The modes encircle the aPf region and the coordinate $x$ follows the modes in the downstream direction. Floating point contacts are at temperatures $T_L$ and $T_R$. (b) The range of equilibration lengths $l_\text{Edge}$ and $l_\text{inner}$ for which both the aPf-2 and aPf-3 interface have the experimentally observed value $\frac{G_Q}{\kappa_0 T} = 0.5\pm0.05$ is shaded. Not all simulated points are shown in favour of clarity.} \label{fig:hydrodynamics}
     \end{figure}

We finally address the question of just how large the equilibration length needs to be in order to produce results consistent with experiment. We use a hydrodynamic model of the edge analogous to those used in  \cite{01_PhysRevB.107.245301,09_PhysRevB.98.115408}. The crucial change is that these both assume good equilibration of all modes with the contacts. We relax this constraint, simulating a periodic system as shown in Fig.~\ref{fig:hydrodynamics}(a), but allow for a finite equilibration length between each mode and the contacts.
Motivated by the experiment, we simulate an edge of width $L=15~\mu\text{m}$ and contact width $W=8~\mu\text{m}$ in the setup shown in Fig.~\ref{fig:hydrodynamics}(a) (we use a two-terminal setup for simplicity, noting that the key physics of decoupling should continue to hold even for a four-terminal measurement). We assume the edge charged mode $\phi$ and edge Majorana $\chi$ both equilibrate well with the contacts and assign a length of equilibration $l_\text{Edge}$ between $\phi$ and $\chi$. For the inner Majorana modes, we assume a length of equilibration $l_\text{inner}$ between them and the edge Majorana mode $\chi$. We assume that there is a similar lengthscale associated with their coupling to the floating contacts -- for concreteness, we take the equilibration length between the inner modes and the contacts to be exactly $l_\text{inner}$. Considering both the cases where the inner Majorana modes propagate downstream and upstream to simulate aPf-3 and aPf-2, we find the results in Figure ~\ref{fig:hydrodynamics}(b). We see that for $l_\text{inner}\gtrsim 10^2~\mu\text{m}$ and for $l_\text{Edge}\lesssim 5~\mu\text{m}$, the conductance matches the experiment to within uncertainty on both interfaces. We estimate that the real system can be well within these ranges, showing agreement between our model and experiment.

This mechanism of the two inner modes decoupling means that the noise profile is also equivalent to the PH-Pf profile. This is in appealing accord with noise measurements that agree with a PH-Pf-like state \cite{01_PhysRevB.107.245301,with_noise_experiment,25_paul2023topological}. Furthermore, any experiment which does not take special care to couple these inner modes to the physical system edge will not be able to differentiate between a geniune PH-Pf state and the mechanism proposed here, which could have either Pf or aPf as the bulk state.

\section{The remaining PH-Pf part} \label{sect:phpf_edge_eq_lengths}
After the complete decoupling of the inner Majorana modes $\psi_{1,2}$, we are left with an edge theory consisting of a downstream $\nu=\frac12$ boson $\phi$ and an upstream edge Majorana mode $\chi$, so $\mathcal{L}_\text{M}^{-1}+\mathcal{L}_c=\mathcal{L}_\text{PH}$. This is equal to the edge of the PH-Pf state, which is believed to be consistent with the experiment when the two modes are well-equilibrated \cite{01_PhysRevB.107.245301,05_Ma_2024,07_PhysRevLett.125.016801,48_Banerjee_2018,50_Dutta_2022,with_noise_experiment,25_paul2023topological}. We expand on our reasoning that the equilibration mechanism mostly suggested in literature would be very weak for realistic parameter values. Then we propose an alternative mechanism which could in principle be strong enough to equilibrate the two modes.

\subsection{Equlibration of $\phi$ and $\chi$ in isolation} \label{sub:phpf_edge_on_its_own}
In terms of edge equilibration, the PH-Pf interface is somewhat special among the quantum Hall liquids -- while most edge theories with multiple modes have operators with scaling dimension $\Delta \leq 2$ connecting the modes, the most relevant operator connecting the modes here is $\hat{\mathcal{O}}=(\partial_x\phi)\chi i\partial_x\chi$ which has $\Delta=3$ \cite{01_PhysRevB.107.245301,05_Ma_2024,07_PhysRevLett.125.016801}. The problem we want to draw attention to here is that the interaction term must be significantly stronger than dimensional analysis would suggest in order to have any hope at equilibration. From Eq.~\ref{eq:best_eq_lenght_general_eq}, we find that 
\begin{equation}
    l_\text{eq}\sim {l_\text{Disorder}}\frac{1}{S^2}\left(\frac{{l}_T}{l_c}\right)^4. \label{eq:ph_pf_eq_length}
\end{equation}
For $l_c$ roughly the pairing scale, we have ${l}_T/l_c\approx 120$. With that, even for relatively short $l_\text{Disorder}\sim l_c$, the equilibration length is large and $S\gg1$ is needed to equilibrate in experiment. For example, to get $l_\text{eq}\approx 5~\mu\text{m}$ (which is needed as shown in Fig.~\ref{main_fig:range_of_validity}) we need $S \gtrsim 10^3$, which is contrary to the expectation that $S\sim 1$. Taking $S\sim 1$ gives $l_\text{eq} \sim 10\text{ m}$, orders of magnitude larger than the experiment size. A width $w_D$ of the impurity potential increases the equilibration rate by $\left(\frac{w_D}{l_c}\right)^2$, but this is generally close to $1$ for a real experiment and not large enough to offset the discrepancy.

So due to the extreme RG-irrelevance of the term $(\partial_x\phi)(\chi i\partial_x\chi)$, it would need a very large pre-factor to be able to equilibrate the system. Furthermore, it suggests that the equilibration length grows extremely rapidly at low temperatures, $l_\text{eq}\sim T^{-4}$. While the first thermal conductance experiment for $\nu=\frac52$ saw an unusually large growth in thermal conductance at low temperatures \cite{48_Banerjee_2018}, later experiments however do not see such a growth \cite{50_Dutta_2022,with_noise_experiment}.
\subsection{Including a trapped impurity}
An alternative, stronger equilibration mechanism may also be possible, as we now sketch. Suppose that the disorder now has an internal degree of freedom, a Majorana zero mode  $\gamma$. Such disorder has been studied previously, albeit in a different setting \cite{03_PhysRevLett.124.126801}. Consider a perturbation of the form
\begin{equation}\label{eq:trapped_majorana_lag_term}
    S \Gamma f(x) (\partial_x\phi) \chi \gamma.
\end{equation}
Here, $S$ and $f(x)$ are dimensionless, while $\Gamma=v/\sqrt{l_c}$ carries the dimensionful constants.
This term corresponds to tunnelling between the Majorana mode $\chi$ and the trapped Majorana impurity $\gamma$ under the assumption that this tunnelling rate has a dependence on the charge density in the bosonic mode. Following Eq.~\ref{eq:best_eq_lenght_general_eq}, we find, assuming $f(x)$ is of finite width $w_D$,
\begin{equation} \label{eq:final_for_trapped_impurity}
    l_\text{eq}=l_T \frac{1}{S^2} \frac{l_\text{Disorder} l_c}{w_D^2}.
\end{equation}
We assume that the density of these impurities, $l_\text{Disorder}$ is similar to the range of influence of each, $w_D$, and take both to be $\sim l_c \sim 100\text{ nm}$, which is experimentally reasonable. Then for $S\sim 3$, we get $l_\text{eq}\sim 2~\mu\text{m}$, which is less than the typical distances between gates in experiment $L\sim 10~\mu\text{m}$ and should be enough to equilibrate in practice, as confirmed in Fig.~\ref{main_fig:range_of_validity} above. While $S$ slightly larger than $1$ is required for this mechanism to equilibrate in the experiment, this value is much lower than $S\sim 10^3$ required without a Majorana impurity. Thus with reasonable parameter choices, a trapped Majorana impurity can equilibrate the two modes well.
It should be noted that this mechanism predicts a different temperature dependence of the equilibration length between the two modes, namely Eq.~\ref{eq:final_for_trapped_impurity} predicts $l_\text{eq}\sim T^{-1}$, a significantly less pronounced scaling than the $l_\text{eq}\sim T^{-4}$ law seen in Sect.~\ref{sub:phpf_edge_on_its_own}. While $T^{-4}$ agrees better with the initial experiment \cite{48_Banerjee_2018}, it cannot explain the lack of growth in the thermal conductance seen in \cite{50_Dutta_2022,with_noise_experiment} as well as $T^{-1}$ can.

Finally, there also exists a coupling between the inner modes and the Majorana impurity, $\psi_{1,2}\gamma$ and also a direct coupling of the impurity to the edge mode, $\gamma\chi$. While the second order tunnelling process $\psi_{1,2}\rightarrow \gamma\rightarrow\chi$ might look like it could equilibrate the modes, it suffers from the same issues regarding momentum mismatch and lack of wavefunction overlap as $\chi\psi_{1,2}$ does. Because of that, it cannot equilibrate $\psi_{1,2}$ with the rest of the edge.

\end{document}